\documentclass[journal]{IEEEtran}
\usepackage{amsmath,amsfonts}
\usepackage{algorithmic}
\usepackage{array}
\usepackage[caption=false,font=normalsize,labelfont=sf,textfont=sf]{subfig}
\usepackage{textcomp}
\usepackage{stfloats}
\usepackage{url}
\usepackage{verbatim}
\usepackage{graphicx}
\usepackage{xcolor}%
\usepackage{textcomp}%
\usepackage{manyfoot}%
\usepackage{booktabs}%
\usepackage{listings}%
\usepackage{cite}
\usepackage{hyperref}

\def\BibTeX{{\rm B\kern-.05em{\sc i\kern-.025em b}\kern-.08em
    T\kern-.1667em\lower.7ex\hbox{E}\kern-.125emX}}
\usepackage{balance}

\begin{document}

\title{RBA-FE: A Robust Brain-Inspired Audio Feature Extractor for Depression Diagnosis}

\author{Yu-Xuan Wu, Ziyan Huang, Bin Hu*, Zhi-Hong Guan
\thanks{This work was supported in part by the National Natural Science Foundation of China under the Excellent Young Scientists Fund 62322311 and by the Guangdong Artificial Intelligence and Digital Economy Laboratory (Guangzhou) under Fund PZL2023ZZ0001. \textit{(*Corresponding author: Bin Hu)}}
\thanks{Y.-X. Wu, Z. Huang, B.~Hu are with the School of Future Technology, South China University of Technology, and also with the Pazhou Lab, Guangzhou 510006, China. Y.-X. Wu and Z. Huang contributed equally to this work. (E-mail: huu@scut.edu.cn).} 
\thanks{Z.-H.~Guan is with the School of Artificial Intelligence and Automation, Huazhong University of Science and Technology, Wuhan 430074, China. (E-mail: zhguan@mail.hust.edu.cn).}
}

%\markboth{Journal of \LaTeX\ Class Files,~Vol.~18, No.~9, September~2020}%
%{How to Use the IEEEtran \LaTeX \ Templates}

\maketitle

\begin{abstract}
This article proposes a robust brain-inspired audio feature extractor (RBA-FE) model for depression diagnosis, using an improved hierarchical network architecture. Most deep learning models achieve state-of-the-art performance for image-based diagnostic tasks, ignoring the counterpart audio features. In order to tailor the noise challenge, RBA-FE leverages six acoustic features extracted from the raw audio, capturing both spatial characteristics and temporal dependencies. This hybrid attribute helps alleviate the precision limitation in audio feature extraction within other learning models like deep residual shrinkage networks. To deal with the noise issues, our model incorporates an improved spiking neuron model, called adaptive rate smooth leaky integrate-and-fire (ARSLIF). The ARSLIF model emulates the mechanism of ``retuning of cellular signal selectivity" in the brain attention systems, which enhances the model robustness against environmental noises in audio data. Experimental results demonstrate that RBA-FE achieves state-of-the-art accuracy on the MODMA dataset, respectively with 0.8750, 0.8974, 0.8750 and 0.8750 in precision, accuracy, recall and F1 score. Extensive experiments on the AVEC2014 and DAIC-WOZ datasets both show enhancements in noise robustness. It is further indicated by comparison that the ARSLIF neuron model suggest the abnormal firing pattern within the feature extraction on depressive audio data, offering brain-inspired interpretability. 
\end{abstract}


\section{Introduction}
\IEEEPARstart{D}{epression}, defined by prolonged periods of low mood, is a global health crisis that significantly impairs the quality of life \cite{1}. According to the World Health Organization, depression affects 3.8\% of the global population and is a leading cause of disability, resulting in a substantial death burden. The COVID-19 pandemic era has accentuated the demand for homestyle mental health self-assessment, necessitating not just cost-effective but also rapid diagnostic methods \cite{2}. Traditional diagnostic approaches, such as psychological surveys or scale questionnaires, require an average of 15-30 minutes \cite{3,4}. More importantly, these approaches require interpretation and analysis by healthcare professionals, thereby delaying the diagnosis. This protracted process often deters individuals from engaging in self-assessment. Yet, modern deep learning methods could be more objective and less time-consuming when used for depression diagnosis. Patients may have to record just a few seconds of audio, sufficient for training a deep learning model. Current deep learning methods predominantly use various datasets like vision \cite{5}, audio \cite{6},\cite{7},\cite{8}, Electroencephalogram (EEG) \cite{9}, or multi-modal data \cite{10,11}. Among the physiological modal data, audio-based diagnosis is more feasible for homestyle self-assessments since the speech data acquisition is much easier than other modalities, e.g., professional X-rays and magnetic resonance imaging (MRI) scans or specialized EEG equipments.
%\smallskip

In the literature, deep learning models have achieved considerable performance when used for audio-diagnosis of depression with exquisitely prepared datasets. For instance, WavDepressionNet predicts levels of depression by analyzing raw voice signals, utilizing both representational and evaluative blocks \cite{6}. Through temporal-spatial self-calibration, it attains an RMSE of 8.20 and MAE of 6.14 on the AVEC 2014 database. Similarly, SpeechFormer++ classifies depression through paralinguistic features, employing unit encoders to emulate inter-speech information and fusing blocks to generate features at various granularities, achieving a WA of 77.1\% on the DAIC-WOZ dataset \cite{7}. Other models like bidirectional long short-term memory (LSTM) and time distributed convolution neural network (CNN) focus on capturing temporal dependencies, with CNN capturing prosodic features and Bi-LSTM focusing on temporal relationships, achieving F1 scores of 0.9870 and 0.9074 for healthy and depressed individuals on the DAIC-WOZ dataset respectively \cite{11}. However, most of these learning models have to be trained on curated datasets, placing limitation on their applicability and robustness. This limitation could be severe particularly in homestyle setups, since recording equipments like cellphones or computers are often non-professional and subject to environmental noises.

For homestyle audio diagnosis, environmental noises significantly degrade the quality of audio capture. According to the United Nations Environment Programme’s 2022 Frontiers Report, noise levels in most global cities exceed acceptable limits, with lower-floor residents frequently suffering from vehicular and human-induced noise \cite{12}. Audio capturing devices like microphones might also bring electrical noises. It has been recognized that the environmental noise and reverberation can degrade the performance of depression diagnosis models \cite{13}. Models such as deep residual shrinkage networks (DRSN) can bee employed for dealing with noises in time-series anomaly detection \cite{14}. However, these methods cannot be directly applied to handle noisy audio samples. For example, key features like jitter \cite{15}, might be thresholded by DRSN, leading to the loss of crucial information. {These observations motivate to design intelligent models with high noise robustness, specifically for handling the task of audio-based depression diagnosis. Pioneering work and recent advance suggest brain-inspired neuronal computations, such as spiking neural networks (SNNs) \cite{16,17} and hybrid neural networks \cite{18,18-1}, which outperform other machine learning models against noises}. Most of such networks employ leaky-integrated-fired (LIF) models that mimic the neural impulse transmission mechanism in the brain. Choosing appropriate thresholds and membrane potential leaks, LIF models can filter out noise to a certain degree\cite{19}. LIF neurons emit discrete action potentials, also called spikes or nerve impulses, for description and transmission, where temporal information can be stored in membrane potentials. Those easily overlooked audio features may accumulate, making the model to release and transmit information in an appropriate event-driven manner.

\begin{figure*}%[ht]  [b]
\centering
\includegraphics[width=0.9\textwidth]{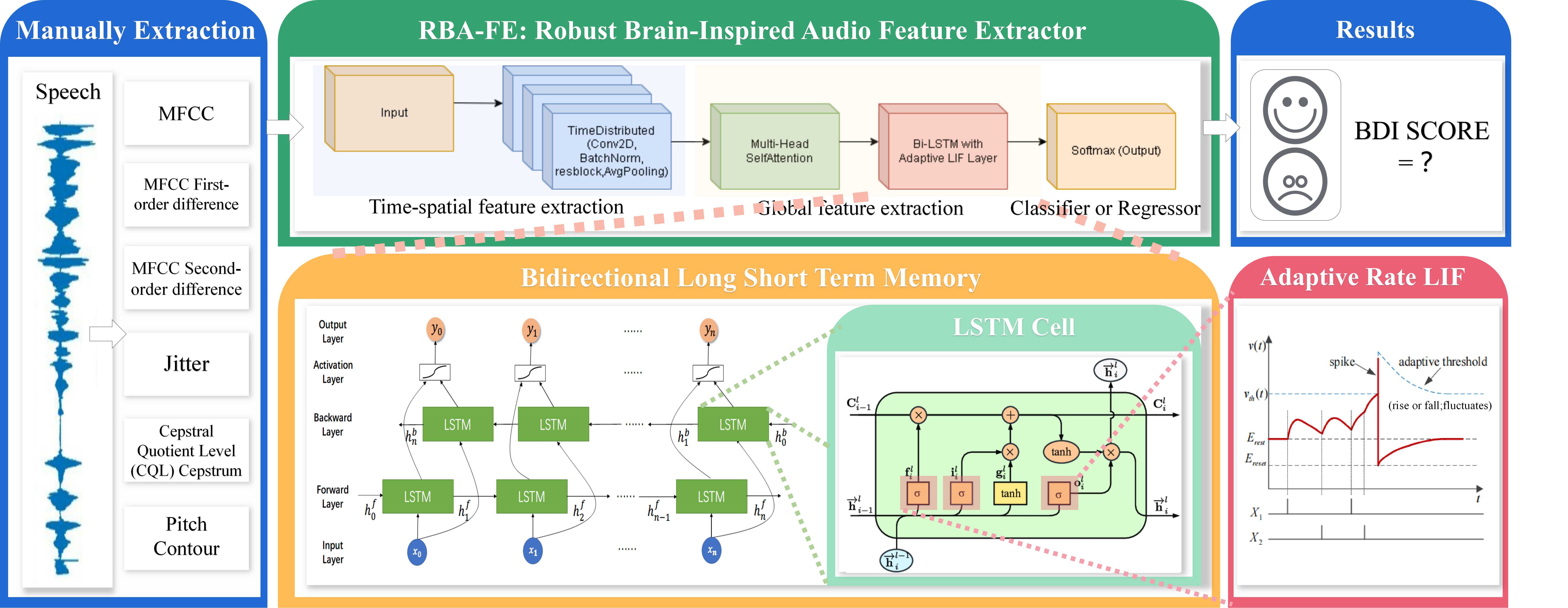} 
\caption{Diagram of the RBA-FE model. Initial inputs are the preprocessed six key features obtained from raw audio signals. These features go through T-CNN, multi-head self-attention modules and Bi-LSTM networks, and then adaptive rate smooth LIF neurons. Self-attention assigns feature weights based on importance, while adaptive rate smooth LIF neurons adjust firing rates according to inputs. Bi-LSTM with adaptive rate smooth LIF neurons improves the network robustness. This adaptive process emphasizes crucial features, revealing patterns for accurate emotion recognition.} \label{fig-1}
\end{figure*}

LIF-based SNN models demonstrate desirable performance for dealing with environmental noises in audio recognition, preserving partial temporal features of the audio. However, one limitation of the standard LIF model is its rigid resetting threshold, failing to adapt to the heterogeneous nature of individual signal points in the time domain \cite{20}. {In order to improve the noise robustness and solve the vanishing gradient issue, we take inspiration from an attention perceptual mechanism of the brain---retuning of cellular signal selectivity \cite{16,21}}. That is, neurons display a higher firing rate in response to preferred stimuli, while non-preferred stimuli, even if potent, elicit no corresponding reaction from the cell. {Following this inspiration, we present an adaptive spiking mechanism to modify the standard LIF model, named adaptive rate smooth leaky integrate-and-fire (ARSLIF)}. In ARSLIF, the neuronal threshold is not constant but adjusts along its target firing rates, facilitating adaptive sensitivity modulation to diverse input signals. In addition, the model incorporates a smoothing operation to reduce gradient noise, improving the training stability of the model.

In this article, a robust brain-inspired audio feature extractor (RBA-FE) is proposed for depression diagnosis, by leveraging ARSLIF neurons, multi-head attention and long short-term memory. The RBA-FE model extracts six acoustic features from raw audio data, which are captured by a local feature extractor for spatial attributes, and then processed by a temporal feature extractor for time dependencies (Fig. \ref{fig-1}). For understanding emotional states\cite{48}, the six acoustic features are Mel Frequency Cepstral Coefficents (MFCC), Pitch, Jitter, CQL Cepstrum. MFCCs are crucial for capturing important spectral characteristics, and their first and second derivatives provide insight into the dynamics in speech \cite{49}. Pitch is another crucial feature, with a reduction in variance and difficulties in pitch recognition being closely associated with the severity of depression, especially in females\cite{50}. Jitter is used to quantify irregularities in vocal frequency, linked to more severe forms of depression \cite{51}.  CQL improves the analysis of harmonic structures, offering better insights into the emotional effects on speech\cite{52}. Noisy data is further filtered out using the ARSLIF model through a classifier. 

At the initial phase, the six acoustical features are manually extracted to reveal various attributes of the audio. Then, RBA-FE utilizes a time distributed convolution neural network to concurrently capture key features from each frame of audio while reducing redundancy through dimensionality reduction. At the second phase, a global feature extractor composed of multi-head attention and bidirectional LSTM refines critical emotional information and further reduces dimensionality. More importantly, this phase is enhanced by the combination of weight allocations in multi-head attention by the aid of the ARSLIF neurons. Each ARSLIF neuron accumulates membrane potentials till threshold and emits spikes; if no subsequent signals are received, its membrane potential leaks, further filtering key information for each time frame. In conjunction with the attention, ARSLIF neurons can further strengthen the weight focus. At the final phase, RBA-FE adopts a softmax layer for classification outputs or a fully connected layer as a regressor for output. Overall, RBA-FE processes the audio depression features by a hierarchical feature extractor. Our main contributions are summarized as follows: 
\begin{enumerate}
\item {To deal with the audio-based depression diagnosis tasks, this article proposes an intelligent method, termed as the robust brain-inspired audio feature extractor (RBA-FE) model}. Different from the visual or physiological modal data, audio-based diagnosis is more feasible for homestyle self-assessments. This trait allows our diagnosis model to be portable.
\item {RBA-FE combines T-CNN, multi-head attention and Bi-LSTM networks, where an adaptive rate smooth leaky integrate-and-fire (ARSLIF) neuron model is presented to tailor the noise challenge in audio. ARSLIF mimics the attention perceptual mechanism of the brain \cite{16,21}, which, together with attention and LSTM, can enhance the noise-filtering capability on audio datasets corrupted by noises}. 
\item By leveraging six acoustic features, RBA-FE can capture both spatial characteristic and temporal dependency, which conventional models like DRSN can not. Applied to the AVEC2014, MODMA, and DAIC-WOZ datasets, RBA-FE outperforms related models in terms of classification accuracy and robustness. 
\end{enumerate}

The rest of this article is organized as follows. Section \ref{sec2} reviews related work on depression audio features and deep learning models. Section \ref{sec3} elaborates on the proposed RBA-FE model and its components. Real data experiments and results are provided in Section \ref{sec4}, including parameter setups and comparative studies. \ref{sec6} concludes the article.
\smallskip

\section{Related Work} \label{sec2}

\subsection{Studies on Depression Audio Features}
Research into diagnosing depression using audio features has gained significant attention over the past few years. Previous studies have established that the neurophysiological changes associated with depression disrupt the coordination of vocal articulation in affected individuals, thereby altering the quality of their speech. These alterations are encoded in the acoustic properties of their speech signals, such as the source, spectrum, prosody, and formant peaks \cite{15}. These observations have led researchers to explore alternative approaches based on audio processing techniques, applicable for practical purposes like automatic depression screening and remote monitoring. Subsequently, various studies have sought clues that reflect mood disorders affecting a subject's audio \cite{23}. Acoustic features of audio signals such as pitch, formants, harmonic-to-noise ratio, shimmer, jitter, speech rate, energy, and glottal features have been employed for the analysis \cite{24,25,26}, of audio in depression sufferers. For example, \cite{24} discovered that rhythm, pitch, and both the intensity and quality of voice can be indicators of an individual's emotional and psychological states. Moreover, it is noted that voices of individuals with depression often exhibit low energy, lack of emotional intensity, and a paucity of rhythmic variations \cite{27}. In this article, to enhance the feature quality, we first manually extracted six features. Then the six features are combined with other automatically extracted features, generating the total model input. This integrated features can improve the interpretability and accuracy of the RBA-FE model. 

\subsection{Deep Learning Models for Audio Depression Diagnosis}
With the advent of deep learning, a great deal of neural network architectures have been developed for audio data analysis. He et al. used CNNs for feature extraction from voice data, with the goal of identifying individuals with or without depression \cite{28}. In recent years, automated methods for diagnosing depression have also gained traction. These methods offer the advantage of providing rapid, low-cost screening for a large patient pool, thus enabling early diagnosis and treatment of depression. They typically focus on how to better extract multi-scale and multi-dimensional voice features associated with depression. For example, Zhao et al. designed one-dimensional and two-dimensional CNN-LSTM networks to learn local and global features for emotion recognition in voice, considering that most models only capture low-level features for emotion classification \cite{30}. Niu et al. developed a WavDepressionNet model that more efficiently extracts emotion-related information by directly modeling raw voice signals\cite{6}. Rejaibi et al.  proposed an EmoAudioNet model that uses one-dimensional convolution (Conv1D) and two-dimensional convolution (Conv2D) to extract depression clues from both time-frequency and spatial perspectives \cite{31}. 

To capture intra- and inter-component hierarchies in speech, Mao et al. designed a unit encoder that can integrate both coarse-grained and fine-grained details. They also constructed a hierarchical backbone, SpeechFormer++, for paralinguistic processing. These models all employ different methods to extract as many useful features as possible from voice data related to depression. These models have been trained and tested based on various publicly available depression voice datasets such as AVEC2014 \cite{40}. However, these datasets typically do not simulate voice data from depression sufferers in noisy environments. Additionally, if artificial intelligence diagnosis models are to be used in an everyday, on-the-go context for self-diagnosis, higher noise robustness requirements will be essential. This further motivates our study of robust learning methods.
\smallskip

\section{RBA-FE: the Robust Brain-inspired Audio Feature Extractor model} \label{sec3}
\begin{figure*}[ht]
  \centering
  \includegraphics[width=0.8\textwidth]{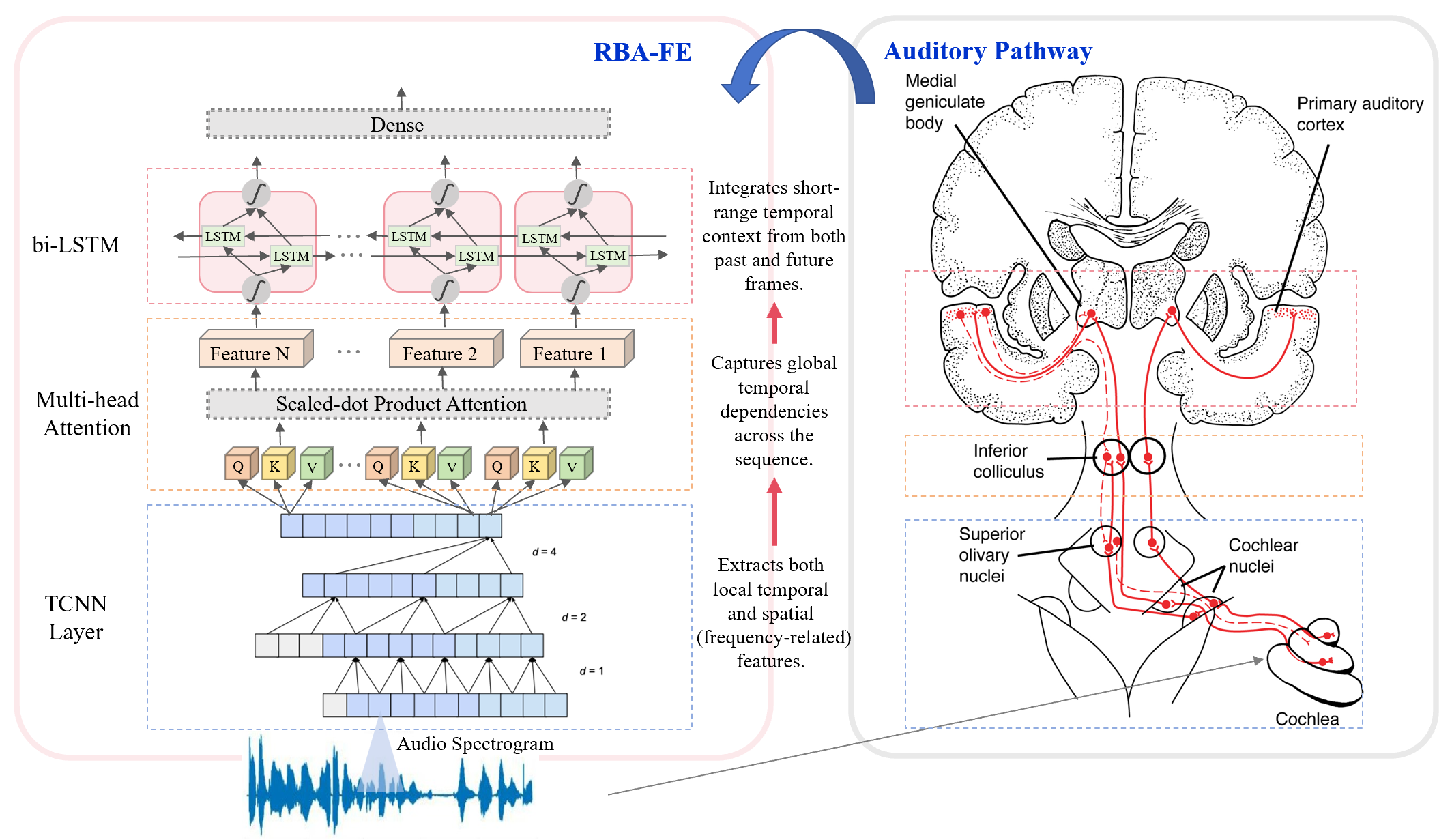}
  \caption{The hierarchical network architecture for feature extraction in the RBA-FE model, with inspiration from the information processing in auditory pathway of human brains. The number $N$ is determined by dividing the total length of the sequence by 1.5 ms. The brain auditory pathway is adapted from \cite{59}.}
  \label{fig-2}
\end{figure*}

\begin{figure}%[ht]
	\centering
	\includegraphics[width=0.49\textwidth]{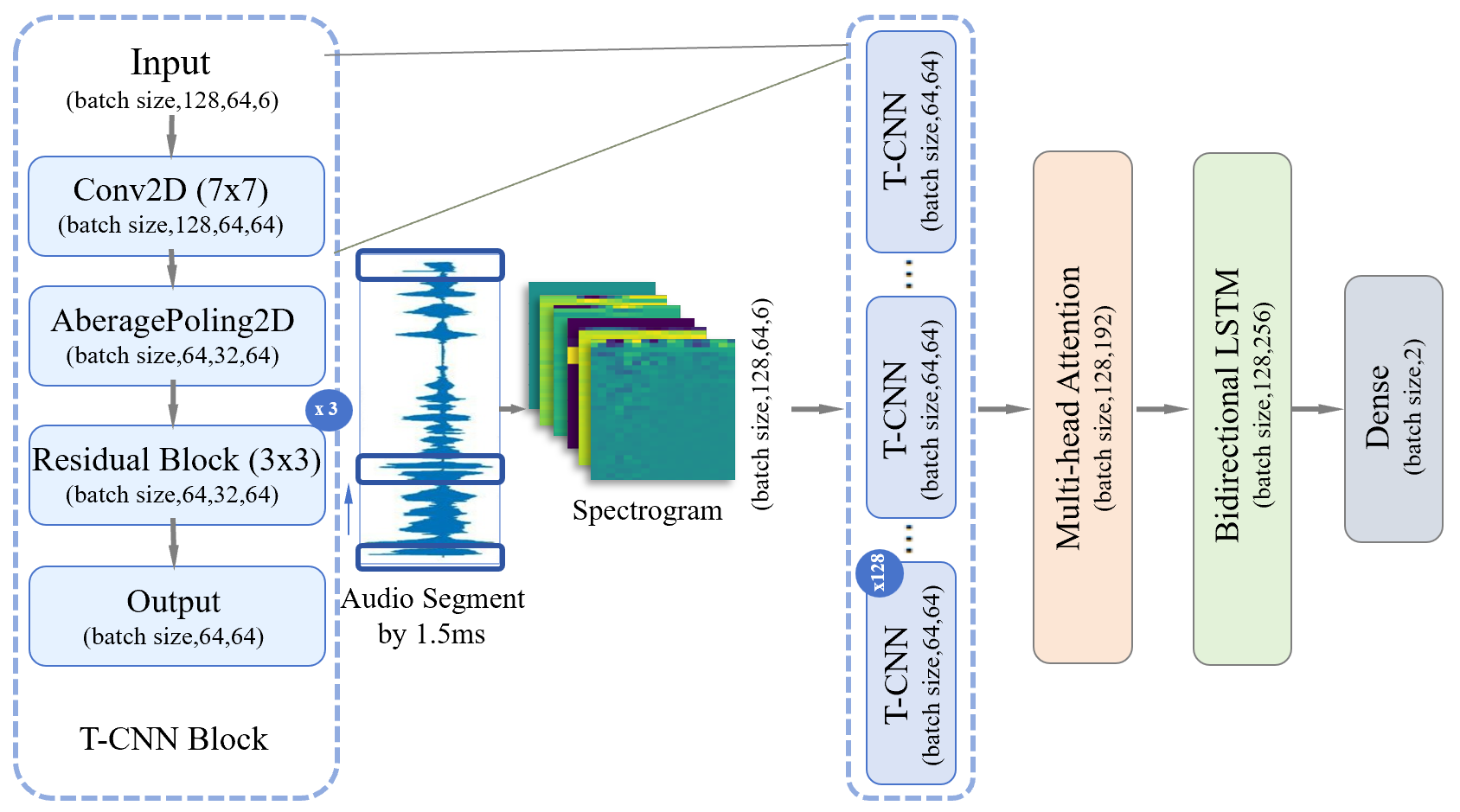}
	\caption{Illustration of data flow and tensor sizes in RBA-FE.}
	\label{fig-2-1}
\end{figure}

This section provides a detailed explanation of the RBA-FE model, in which the hierarchical network architecture is inspired by the auditory processing pathway in the human brain (Fig. \ref{fig-2}). Researches on the auditory pathway demonstrates that sound processing in the brain progresses through multiple hierarchical stages, from basic acoustic feature extraction in early auditory regions, such as the cochlear nuclei and superior olivary complex to complex temporal processing in the cortex \cite{58,62}. It is further shown that these stages correlate with specific layers of deep neural networks (DNNs) for audio processing, supporting the use of a hierarchical architecture for robust feature extraction \cite{59,60}. By this brain inspiration, RBA-FE utilzies T-CNN for early feature processing, multi-head attention for selective focus on relevant parts of the audio spectrogram, Bi-LSTM for integrating information across frames, and a softmax layer for classification or a fully connected layer for regression (Fig. \ref{fig-2-1}). 

%这个还没有修改

\subsection{Foundational components of the RBA-FE model}
For audio-based depression diagnosis, RBA-FE integrates temporal convolutional neural networks (T-CNN), multi-head attention, and bidirectional long short-term memory networks (Bi-LSTM), leveraging the unique strengths of each element to tackle complex diagnostic challenges.

\subsubsection{Temporal Convolutional Neural Network}
The RBA-FE starts with a T-CNN. To deal with temporal data analysis\cite{32}, T-CNN excels in local feature extraction, {and its cascade of convolutional and pooling layers progressively unveils signal intricacies}, capturing rhythm, intensity, and pitch within speech signals. {As shown in Fig. \ref{fig-2-1}, the audio data is first pre-processed by manual feature extraction at a sampling frequency of 1024Hz, utilizing a window size of 1024 without overlap. Segmentation of the audio is performed into intervals of 1.5ms, while the segment shorter than 1.5ms is repeated to approximate the interval length, discarding any excess. Then, the segmented data undergoes convolution with $7\times 7$ filters.} This is followed by batch normalization, ReLU activation, and average pooling. The data then undergoes further processing with residual blocks of $3\times 3$ filters for enhanced feature extraction. The output of the T-CNN, concatenated along the temporal dimension, serves as the input for the subsequent attention layer. 

Employing two-dimensional convolutional layers and dilated convolutions, a T-CNN layer increases its receptive field to capture broader temporal relationships within the depression speech feature map, identifying subtle vocal modulations and global audio events. This layered structure resembles the processing stages of the early auditory pathway, where each successive layer captures increasingly complex auditory features, progressing from simple spectral patterns to more sophisticated temporal structures in a manner akin to the transformation from cochlear nuclei to higher auditory cortex regions. T-CNN makes the foundation for subsequent network layers, much like the early auditory processing stages provide foundational acoustic representations for higher-order processing in the brain \cite{58,59}.

\subsubsection{Multi-Head Attention} 

RBA-FE utilizes multi-head attention to compute similarities across all points in the sequence, endowing the model with global dependencies across distant temporal points. Unlike other studies that place attention after LSTM and before fully connected layers, we place the attention after TCNN and before the Bi-LSTM. The adopted attention mechanism intends to amplify sparse features within the speech feature map, enabling the subsequent Bi-LSTM to converge rapidly toward key depressive speech characteristics.
	
More formally, the attention mechanism employs eight parallel attention ``heads", learning different representational subspaces of the input data, which aligns with the selective processing seen in midbrain structures, such as the inferior colliculus, where relevant auditory information is filtered and enhanced. This setup allows the attention mechanism to focus on audio features critical to the diagnosis, such as emotional fluctuations and nuanced speech patterns, while discarding irrelevant background information. By capturing diverse aspects of speech signals, including intonation, rhythm, and semantic nuances, the attention layer mimics the auditory system's capacity to selectively process specific auditory cues. A dropout layer with a 75\% probability is included to prevent overfitting, enhancing the model’s interpretive capabilities.The attention mechanism in RBA-FE enhances the feature map by focusing on key auditory cues. This selective focus refines the extracted features, making them more interpretable and robust for subsequent processing by the Bi-LSTM.

\subsubsection{Bidirectional Long Short-Term Memory Network} 

To capture both immediate and extended temporal dependencies in speech, the RBA-FE model incorporates a Bi-LSTM layer following the attention mechanism \cite{34}. Unlike standard LSTMs, which retain information only from past audio data, Bi-LSTMs capture both past and future context by employing two independent LSTM layers (forward and backward) \cite{45, 46}. This placement after the attention layer allows Bi-LSTM to filter out noisy long-distance correlations introduced by the attention mechanism, focusing the network on short-range temporal dependencies that are more relevant to identifying depression-related features in speech.
	
The forward LSTM processes data from past to future, while the backward LSTM processes data from future to past. When used for time-related audio signal segmentation, Bi-LSTM enables comprehensive understanding of temporal sequences by capturing both past and future information with complex(related or different) emotional or behavioral patterns through the two LSTM units. It is expected that the combination of T-CNN, Attention and Bi-LSTM offers more preciseness in complex spatio-temporal feature extraction with emotional audio conditions. More importantly, LSTM in our model is improved with one single layer of adaptive rate-smoothing Leaky Integrate-and-Fire (ARSLIF) neurons to assist in the task of noise filtering, emulating the nerve impulse responses. The outputs of the two LSTM layers are combined by summation, followed by a dropout layer with a 50\% probability to mitigate overfitting. This bidirectional processing mechanism mirrors higher-order auditory processing in the brain, where signals are integrated across multiple time frames to provide a dynamic and nuanced understanding of speech patterns \cite{16}. 

\subsection{Adaptive Rate Smooth Leaky Integrate-and-Fire Model}
An improved spiking neuron model is presented to deal with the environmental noise and reverberation in voice recordings. It is termed as adaptive rate smooth leaky integrate-and-fire model (ARSLIF), beyond the standard leaky integrate-and-fire (LIF) model. ARSLIF has the capacity of noise filtering akin to the brain, meanwhile ARSLIF also renders the flexibility to preserve salient temporal features in audio data, as will be demonstrated in the experiment section.

\subsubsection{LIF}
The leaky integrate-and-fire (LIF) neuron model is a simplistic, effective neuronal model, capturing the electrophysiological characteristics of neurons in the brain \cite{16,17}. The model draws inspiration from the biological neurons' ability to integrate incoming signals and activate upon exceeding a certain threshold. In biological systems, the membrane potential \( V_{\rm m}(t)\) gradually accumulates in response to the input signal \( I(t) \), triggering an action potential or ``spike" upon reaching a certain threshold, meanwhile the neuronal potential will be reset to a predefined state.

The standard LIF model is given by \cite{16,17}:
\begin{equation}\label{eq-1}
\tau\frac{dV_{\rm m}(t)}{dt} = -V_{\rm m}(t) + RI(t)
\end{equation}
where \(\tau>0\) is the time constant, \(R>0\) is the membrane resistance constant, \(I(t)\) is the input current, and \(V_{\text {rest}}\) is the resting potential. {If $V_{\rm m}$ crosses a potential threshold, then the neuron fires a `spike'; meanwhile the membrane potential is reset to a predefined potential $V_{\text{reset}}$. Let $V_{\text {th}}^0$ be the potential threshold, which usually takes a constant value in the standard LIF model \cite{16,17}. Let $S_i(t)$ be the firing state at time $t$. It follows that}
\begin{equation}\label{eq-2}
	S_i(t) = \begin{cases} 
		1, & \text{if } V_{\rm m} \geq V_{\text{th}}^0 \\
		0, & \text{otherwise}
	\end{cases}
\end{equation}
\begin{equation*}
V_{\rm m} = V_{\text {reset}}, \text { if } V_{\rm m} \geq V_{\text {th}}^0
\end{equation*}
By simulating the spiking mechanism described by eqs. (1) and (2), the standard LIF model is capable of noise filtering and feature extraction when used for machine learning \cite{20}. 
\medskip

\begin{figure}[bp]
	\centering
	\includegraphics[width=0.49\textwidth]{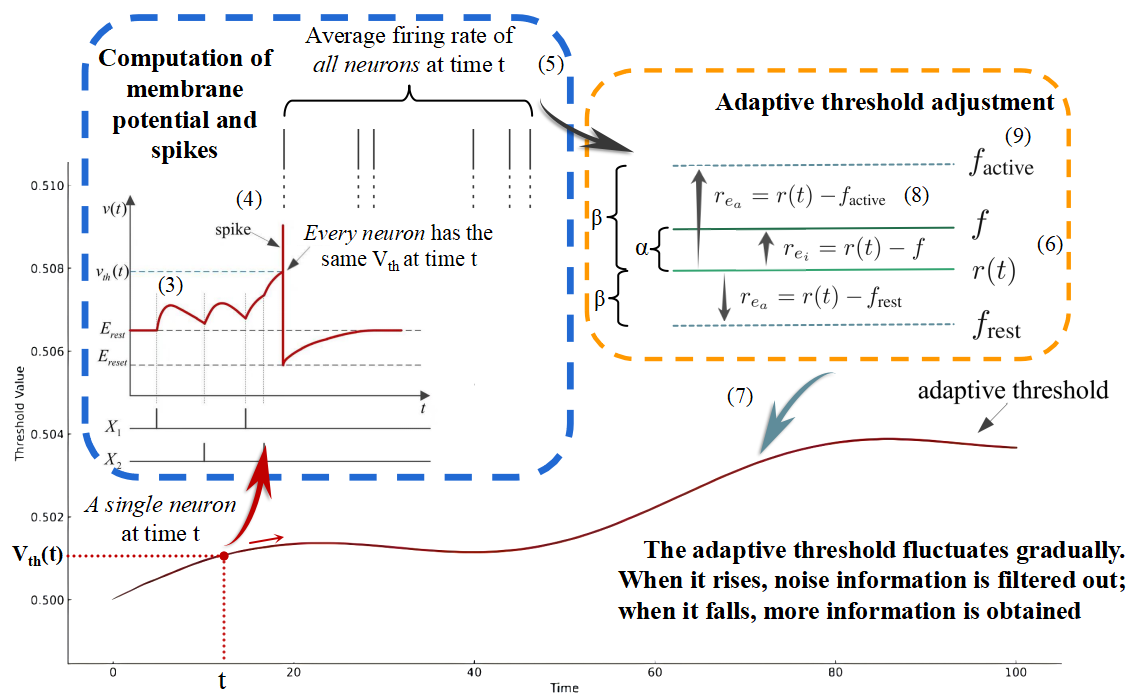}
	\caption{Illustration of the calculation of ARSLIF based on equations (3)-(8), explaining how the adaptive threshold work.}
	\label{fig-1-1}
\end{figure}

\subsubsection{ARSLIF}
The design of the ARSLIF model draws inspiration from the brain. Neurons in the brain utilize action potential (spike) for communication and have a temporally-correlated internal state known as the membrane potential \cite{16}. While the standard LIF model is a simplified neural model that mimics the electro-signaling mechanisms of brain neurons by setting a fixed threshold and membrane leakiness, it may not fully capture the heterogeneity of signals. An adaptive mechanism is introduced to allow the model to adapt more flexibly to different signal patterns, see Fig. \ref{fig-1-1} for an illustration. 

The updating process of the ARSLIF neuron model in training is presented as follow.
\begin{enumerate}
    \item \textbf{Initialization and weight construction:}
    Through the constructor, the ARSLIF model sets multiple key parameters:
\begin{itemize}
    \item \textit{`$f$=target\_rate'}: The desired average firing rate for neurons.
    \item \textit{`$f_{\rm {adapt}}$=target\_rate\_adapt'}: The adaptive target firing rate.     
    \item\textit{`$f_{\text {active}}$=target\_rate\_active'}: The target firing rate used when input potential is above the threshold.
    \item\textit{`$f_{\text {rest}}$=target\_rate\_rest’}: The target firing rate used when input potential is below the threshold.
    \item \textit{`$\tau_{\text{adapt}}$=tau\_adapt'}: The time constant, controlling the speed of threshold adjustment.
\end{itemize}
During the construction phase, the model initializes two critical state variables:
\begin{itemize}
    \item $V_{\text{mem}}$: Membrane potential.
    \item $V_{\text{th}}$: Activation threshold, representing the adaptive potential threshold for spike activation.
\end{itemize}

\item \textbf{Computation of membrane potential and spikes:} The spiking mechanism of ARSLIF is similar to the LIF model:
\begin{itemize}
    \item \textit{Membrane potential update:} A discretization of differential equation (\ref{eq-1}) is used to simulate the membrane potential's leakiness and accumulation of inputs.
    \begin{equation}\label{eq-3}
    V_{\text{mem\_new}} = V_{\text{mem}} - \frac{dt}{\tau_{\text{mem}}} \cdot (V_{\text{mem}} - R \cdot \text{inputs})
    \end{equation}
    where $V_{\text{mem\_new}}$ is the updated new potential.
    \item \textit{Spike generation: }A spike is generated when the membrane potential reaches the threshold, and the membrane potential is reset. Differing from (\ref{eq-2}), the potential threshold $V_{\text{th}}$ is not a constant. One has
\begin{equation*}%\label{eq-2-1}
   S_i(t)= \begin{cases} 
      1, & \text{if } V_{\text{mem\_new}} \geq V_{\text{th}} \\
      0, & \text{otherwise}
    \end{cases}
\end{equation*}
Using the constant threshold $V_{\text{th}}$, the average firing rate of all neurons can be defined by
\begin{equation}\label{eq-4}
	r(t)=\frac{1}{n} \sum_{i=1}^n S_i(t)
\end{equation}
\end{itemize}

\item \textbf{Adaptive threshold adjustment:}
The activation threshold $V_{\text{th}} $ is adaptively adjusted by calculating the error with the target firing rate related to $r(t)$.

\begin{itemize}
    \item \textit{Rate error:} The error calculation takes into account a combination of the desired firing rate and the adaptive target firing rate, satisfying 
\begin{equation}\label{eq-5}
r_e = \alpha \cdot r_{e_i} + \beta \cdot r_{e_a} 
\end{equation} 
where \( r_e = \text{rate\_error} \), \( r_{e_i} = \text{rate\_error\_initial} \), \( r_{e_a} = \text{rate\_error\_adaptive} \), and $\alpha$ and $\beta$ can be chosen to meet different requirements. 

\item \textit{Adaptive law:} The activation threshold is updated by
\begin{equation}\label{eq-6}
V_{\text{th\_new}} = V_{\text{th}} + \frac{dt}{\tau_{\text{adapt}}} \cdot r_e
\end{equation}
where $V_{\text{th\_new}}$ is the updated new threshold.
\end{itemize}
\end{enumerate}

In the equation (5), $r_{e_i}$ and $r_{e_a}$ are defined respectively as 
\begin{equation}
	r_{e_i}=r(t)-f, \quad r_{e_a}=r(t)-f_{\text {adapt}}(t) 
		\label{eq-7}
\end{equation}
where 
\begin{equation}
f_{\rm {adapt}}(t)= \begin{cases}f_{\text{active}},  & \text {if} \mid R \cdot \text{inputs}\mid\ge V_{\text{th}} \\ f_{\text {rest }}, & \text {otherwise }\end{cases}
	\label{eq-8}
\end{equation}
The term $\frac{dt}{\tau_{\text{adapt}}} r_e$ represents the firing rate error with respect to the desired firing rate and the adaptive firing rate, at time scale $\tau_{\text{adapt}}$. In this case, the equation (\ref{eq-6}) suggests that every spike would cause an adaptation of the threshold $V_{\mathrm{th}}$. Altogether, the equations (\ref{eq-5})-(\ref{eq-8}) constitute the adaptive threshold of spiking neurons via the ARSLIF model, differing from the constant $V_{\text{th}}^0$ within the standard LIF (\ref{eq-2}). The above adaptive adjustment mechanism is proposed to enhance the model adaptability, since it enables our whole model to adjust the activity level of neurons in response to varying input conditions.

\subsection{A theoretical analysis of ARSLIF}
\subsubsection{Influence of parameters $\alpha$ and $\beta$} 
$\beta$ tunes the error between $f_{\rm {adapt}}(t)$ and $r(t)$ and $\alpha$ adjusts the error between $f$ and $r(t)$.

Eq. (\ref{eq-8}) suggests that, if the input signal $\rm inputs$ reaches the activation threshold $V_{\rm th}$, indicating that all neurons may be biased towards activation, then the adaptive target firing rate is set to $f_{\text {active}}$ (usually 0.99); otherwise, it is set to $f_{\text {rest }}$ (usually 0.01). This mechanism adjusts the threshold in response to continuous high-rate firing signals, making it easier for the model to reach the activation threshold and thus forming a positive feedback loop that drives the firing rate towards $f_{\text {active}}$. Conversely, if the rate is below the threshold, it drives towards $f_{\text {rest }}$.  $\beta$ then governs the threshold increase process. However, too fast or extreme adaption can potentially disrupt the steady-state of the system. To address this issue, we have introduced the target firing rate $f$ and use $\alpha$ to control the error between it and $r(t)$, in order to limit the final firing rate to oscillate around the target rate $f$.

Combining Eqs. (\ref{eq-5})-(\ref{eq-8}), one has 
\begin{equation*}
	r_e(t)=\alpha \cdot(r(t)-f)+\beta \cdot\left(r(t)-f_{\text {adapt }}(t)\right)
\end{equation*}
From equation (\ref{eq-6}), it follows 
\begin{equation} \label{eq-11}
\frac{d V_{\mathrm{th}}(t)}{d t}=\frac{1}{\tau_{\text {adapt }}}\left(\alpha \cdot(r(t)-f)+\beta \cdot\left(r(t)-f_{\text {adapt }}(t)\right)\right)
\end{equation}
If $V_{\rm th}(t)$ converges to a steady state, the firing rate reaches zero. Letting 
\begin{equation}
0=\alpha \cdot(r(t)-f)+\beta \cdot\left(r(t)-f_{\text {adapt }}(t)\right)
\end{equation}
gives the equilibrium firing rate
\begin{equation}
r(t) = \frac{\alpha}{\alpha + \beta} f + \frac{\beta}{\alpha + \beta} f_{\text{adapt}}(t)
\label{eq-13}
\end{equation}
Here, $\frac{\alpha }{\alpha+\beta}$ is termed as a target rate error ratio, representing the weight of target firing rate $f$ on stability of the ARSLIF model. 

This equilibrium firing rate is a weighted average of the fixed target rate $f$ and the adaptive rate $f_{\text{adapt}}(t)$. Notice that, with a large $\alpha$ and a small $\beta$, $f$ has a higher weight while $f_{\text{adapt}}$ has a lower weight. Since the fixed target firing rate $f$ is constant with respect to the inputs, a higher weight would make $r(t)$ more stable and less sensitive to input fluctuations. By equation (\ref{eq-8}), the adaptive firing rate $f_{\text{adapt}}$ depends on whether the input exceeds the threshold, then a lower weight for it slows $r(t)$ to external changes. In this context, there is a possible balance between stability and adaptability. 
	\begin{itemize}
		\item With a large $\alpha$ and small $\beta$, the neuron’s firing is more stable but adapts more slowly.
		\item With a large $\beta$ and small $\alpha$, the neuron adapts quickly but becomes more sensitive to fluctuations, degrading stability.
\end{itemize}

%\textcolor{brown}{For stability, we want:}

%\textcolor{brown}{$$
%|r(t)-f|<\epsilon \quad \text { and } \quad\left|r(t)-f_{\text {adapt }}(t)\right|<\epsilon$$}
%\textcolor{brown}{Small error $\epsilon$ means the firing rate should stay close to both the initial and adaptive target rates.}
%\medskip

\subsubsection{Advantage of ARSLIF over LIF under noise}

Let the input signal \( I(t) \) be a combination of the actual signal \( I_{\text{signal}}(t) \) and an external noise term \( \eta(t) \). Without loss of generality, \( \eta(t) \) is a Gaussian white noise satisfying 
$$
\langle \eta(t) \rangle = 0, \quad \langle \eta(t) \eta(t') \rangle = \sigma^2 \delta(t - t')
$$
where $\sigma^2$ is the noise variance.
\smallskip

For the standard LIF model, the noise-induced spiking power with noise strength of duration \( T \) can be approximated as:
\begin{equation*}
P_{\text{noise}} = \int_0^T \eta^2(t) dt
\end{equation*}
Since \( \eta(t) \) is Gaussian white noise, its power is proportional to \( \sigma^2 \), leading to the approximation:
\begin{equation*}
P_{\text{noise}} \approx \sigma^2 T
\end{equation*}
This formula indicates that noise directly increases the firing rate because of the fixed threshold, reducing the signal to noise ratio (SNR).
\smallskip

For the ARSLIF model, the adaptive threshold \( V_{th}(t) \) increases with a raising noise by (\ref{eq-11}), with the goal of filtering noise-induced spikes. So the power of noise-induced spiking in ARSLIF is reduced because the threshold increases with noise:
\begin{equation}
	P_{\text{noise}}^{\text{ARSLIF}} = \int_0^T \left( \eta(t) - \frac{1}{\tau_{\text{adapt}}} \triangle V_{th}(t) \right)^2 dt
\end{equation}
Since \( \triangle V_{th}(t) \) increases with noise, this term reduces the overall power of noise-induced spiking, yielding an improved SNR. More specifically, one has
\begin{equation}			
	SNR_{\text{ARSLIF}} = \frac{P_{\text{signal}}}{P_{\text{noise}}^{\text{ARSLIF}}}>\frac{P_{\text{signal}}}{P_{\text{noise}}} =SNR_{\text{LIF}}
\end{equation}
Therefore, using the adaptive threshold mechanism, SNR in ARSLIF is improved compared to the standard LIF.

\subsection{Embedding ARSLIF in the model}

The ARSLIF neuron model is used to replace the conventional sigmoid gate activation function in Bi-LSTM, and the embedding purpose is to cope with the gate overfitting caused by sensitivity to input data noise, as well the noise amplification due to recursion \cite{35, 36}. The ARSLIF neuron model is a biologically plausible mathematical model, reflecting the dynamic behavior of real neurons. By replacing the sigmoid gate activation function in LSTM with ARSLIF, the model can elicit superior neuron-like responses to external signals, particularly in noisy conditions.

% \begin{figure}[htbp]  
	%   \centering
	%   \includegraphics[width=0.4\textwidth]{fig3.png}
	%   \caption{Visualization of the noise robustness of the ARSLIF model. \textit{Left}: Visualizes a 128-dimensional vector's transformation using a Sigmoid function. Lighter shades depict values closer to 1, while darker shades indicate values closer to 0. \textit{Right}: Depicts the vector's change through the original threshold-based ARSLIF function. Lighter shades show spike firing (value of 1), while darker shades signify no spike firing (value of 0).}
	%   \label{fig-3}
	% \end{figure}

\begin{figure*} [htbp]
	\renewcommand\arraystretch{1.3}
	\centering
	\subfloat[]{\includegraphics[width=2.3in]{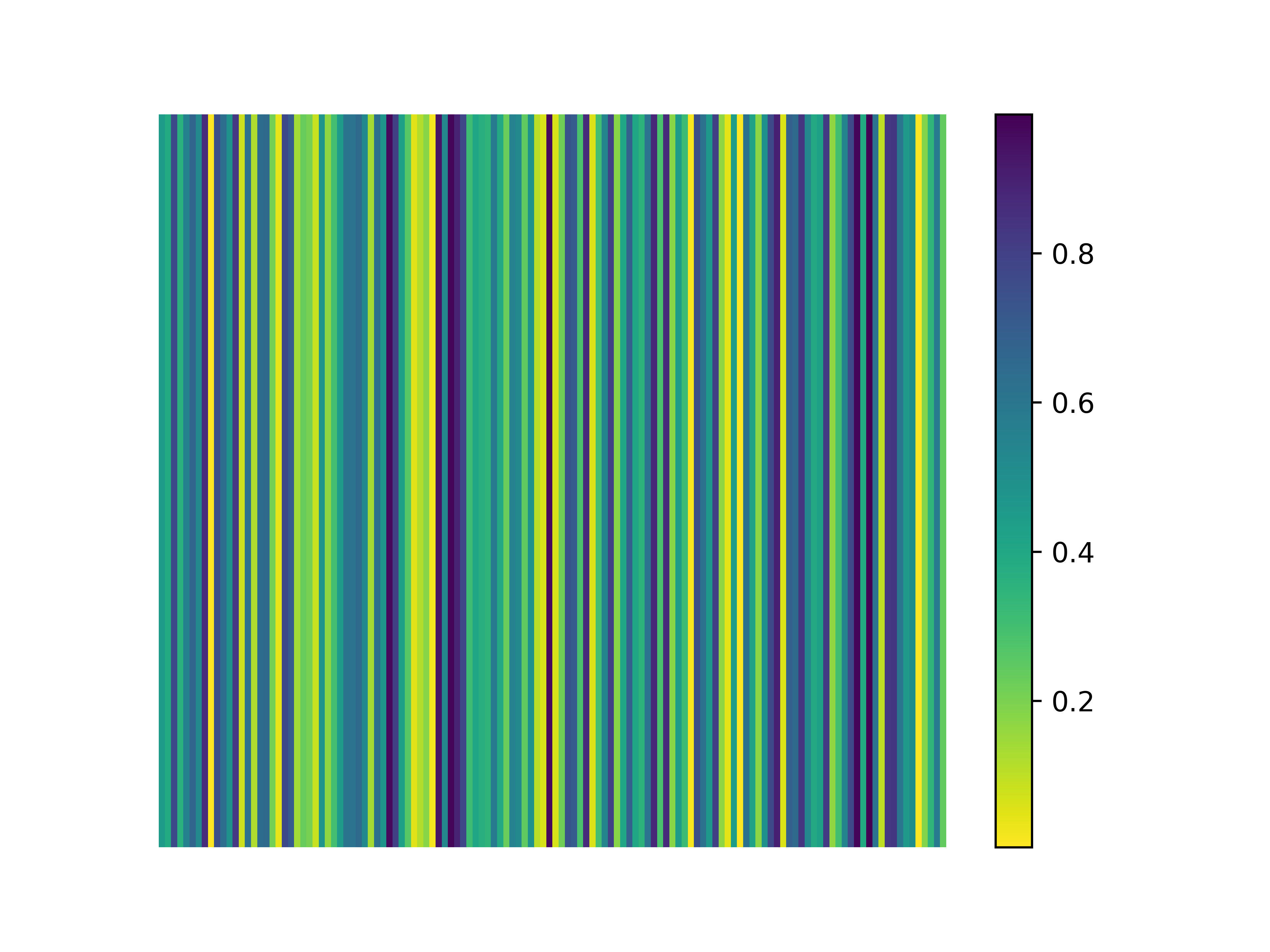}%
		\label{fig_3-1}}
	\hfil
	\subfloat[]{\includegraphics[width=2.3in]{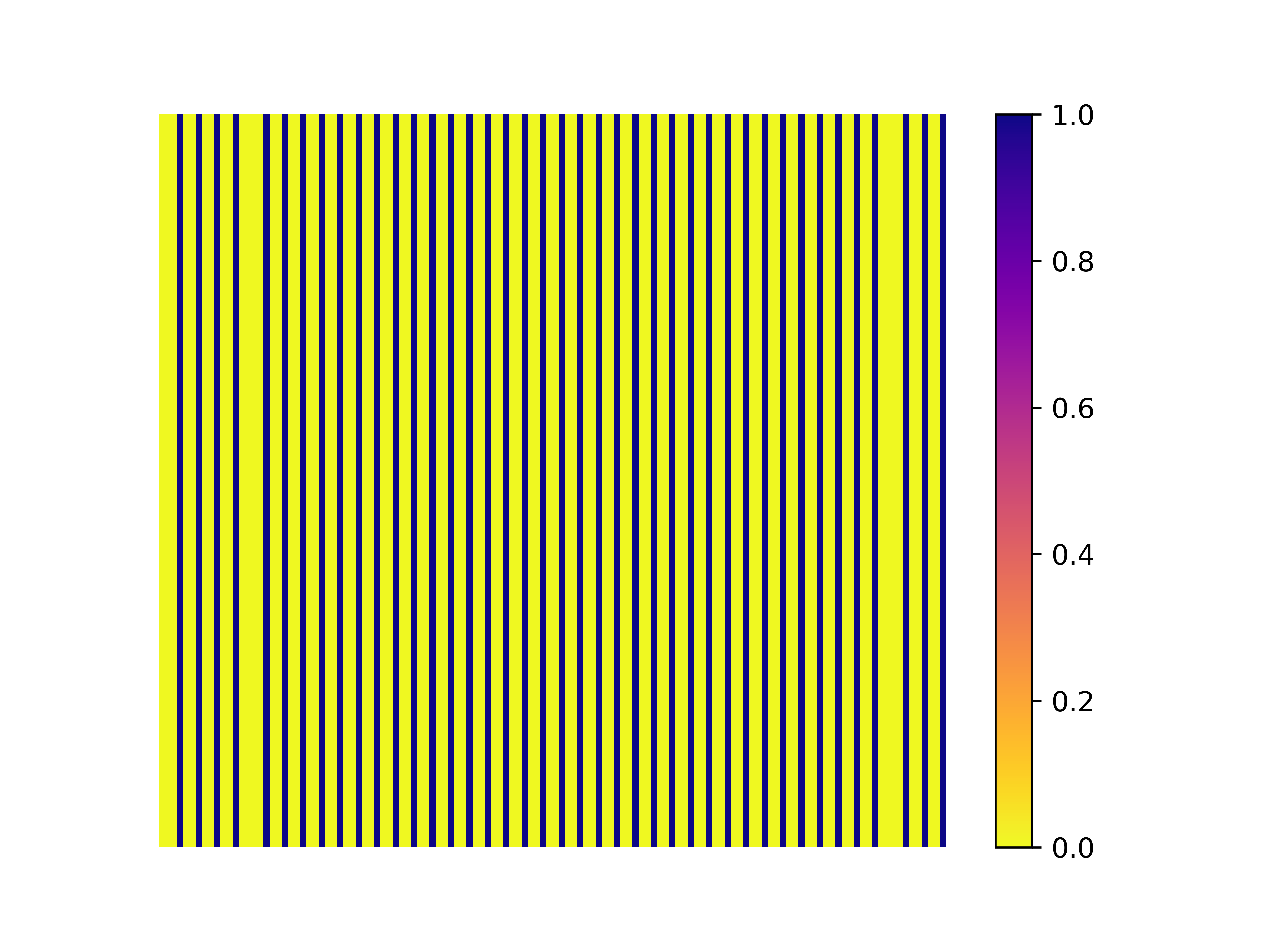}%
		\label{fig_3-2}}
	\hfil
	\subfloat[]{\includegraphics[width=2.3in]{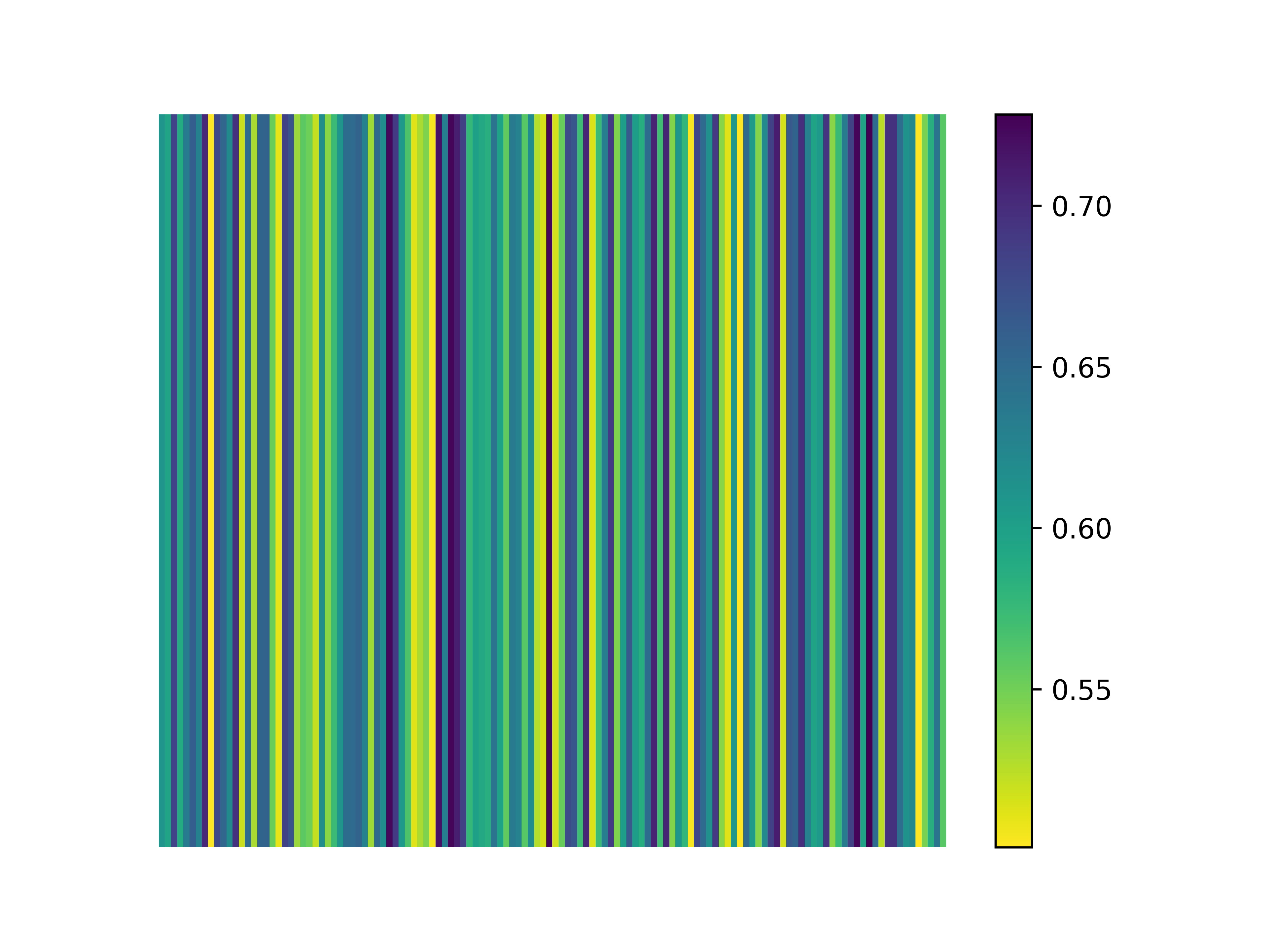}%
		\label{fig_3-3}} \vspace{-0.17in}
	\hfil
	\subfloat[]{\includegraphics[width=2.3in]{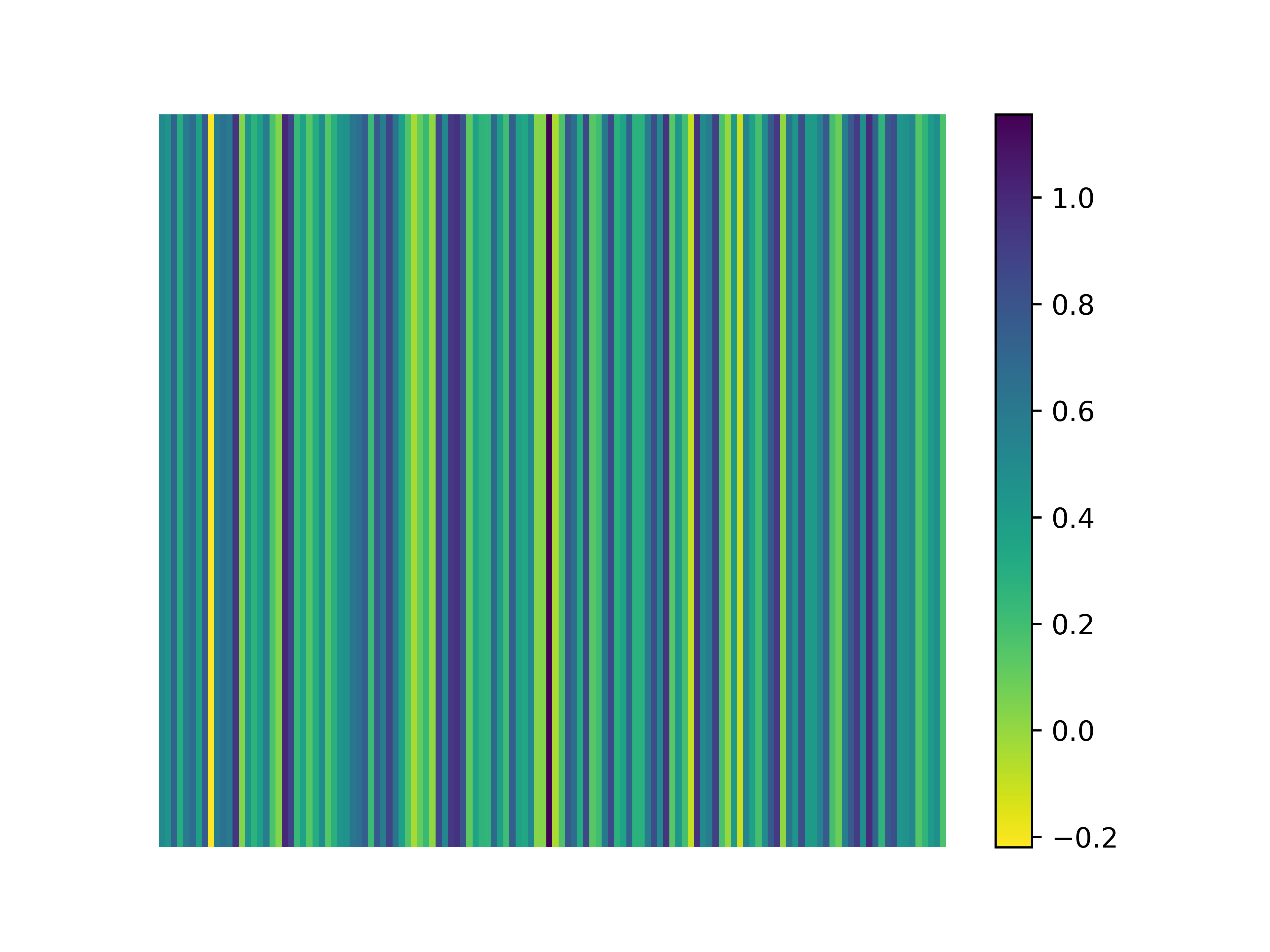}%
		\label{fig_3-4}}
	\hfil
	\subfloat[]{\includegraphics[width=2.3in]{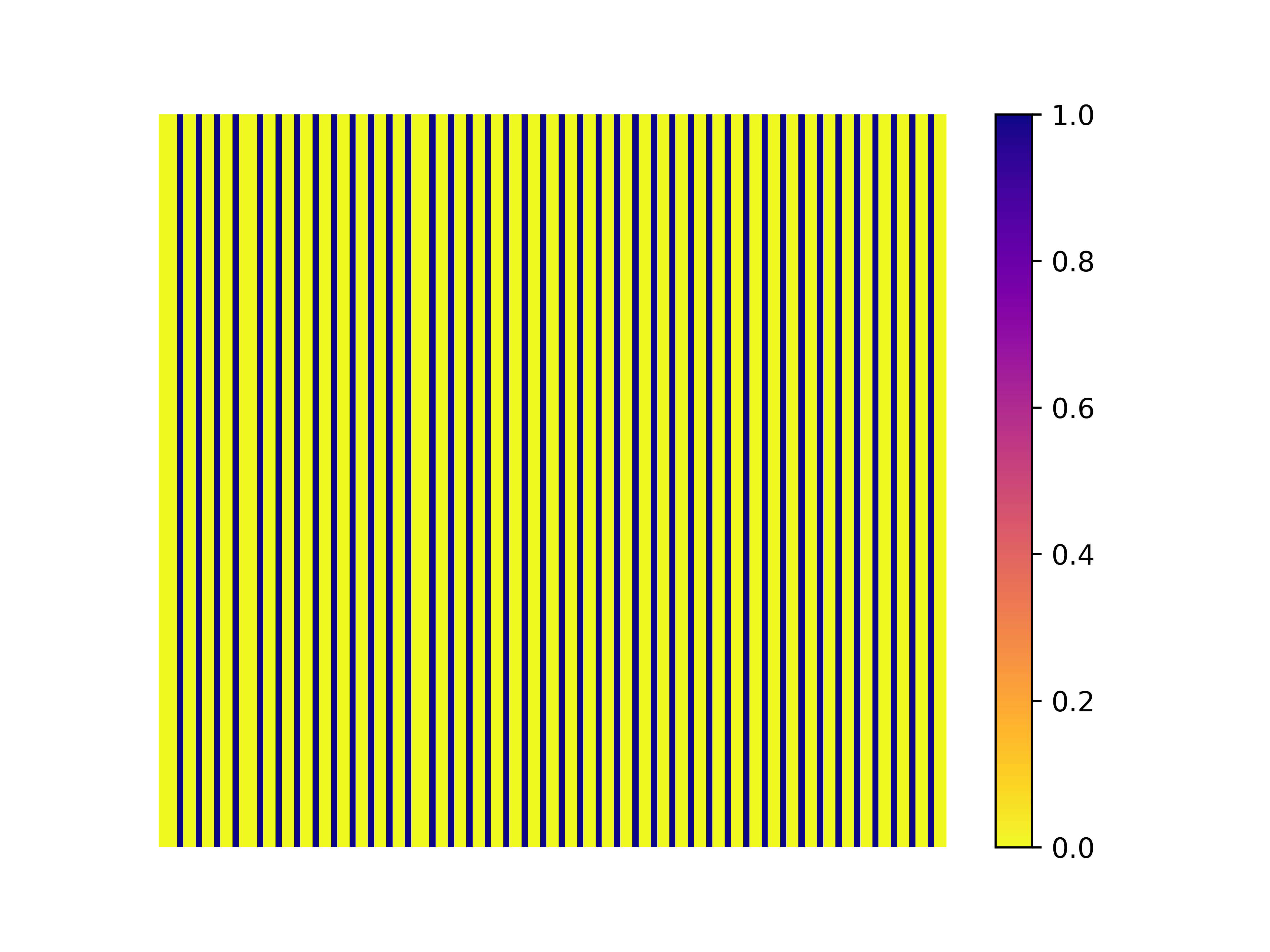}%
		\label{fig_3-5}}
	\hfil
	\subfloat[]{\includegraphics[width=2.3in]{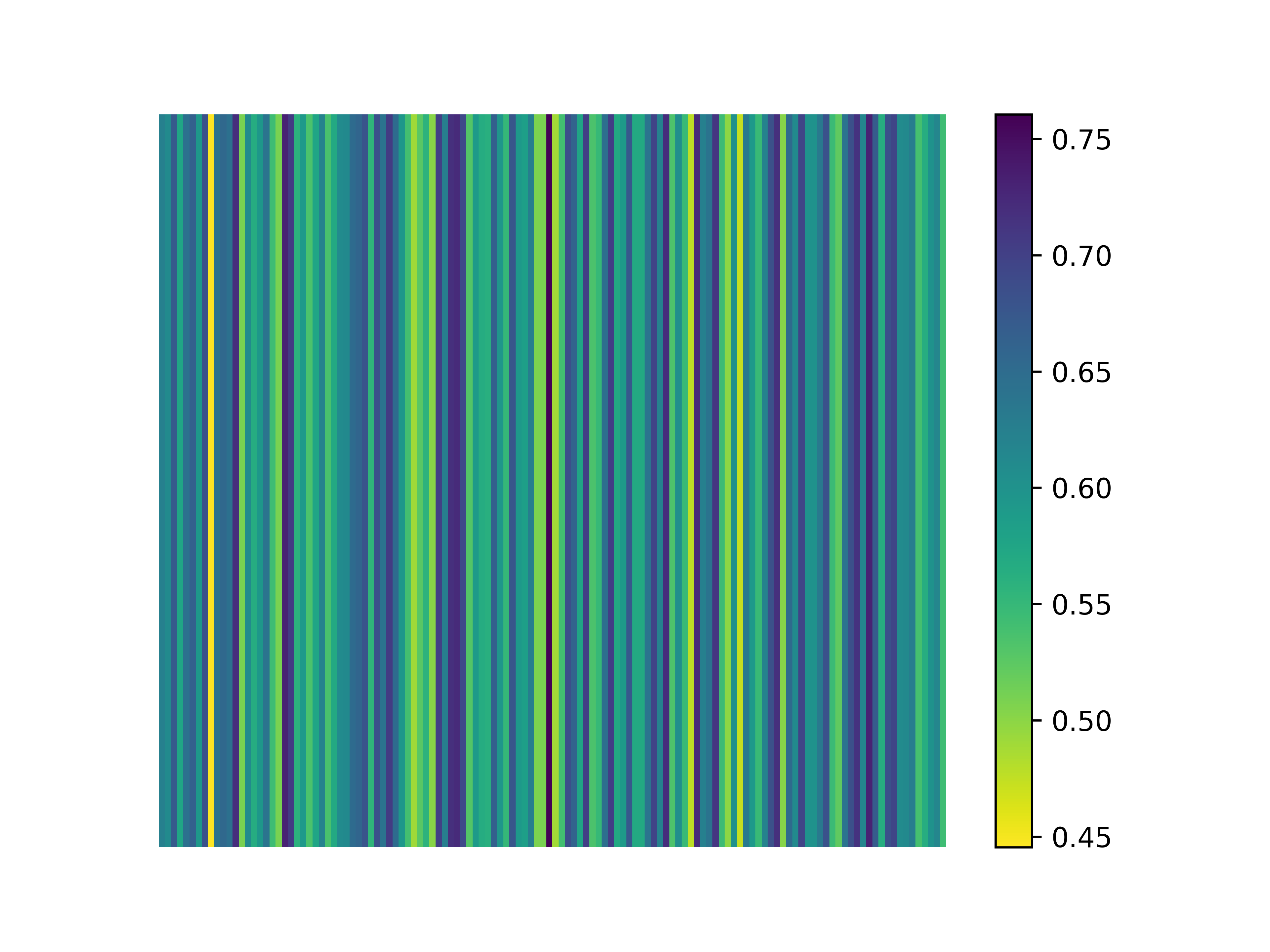}%
		\label{fig_3-6}}
	\caption{Comparative Analysis of ARSLIF and Sigmoid Responses to Clean and Noisy Signals. (a) is the unperturbed input signal, (b) represents ARSLIF response to clean signal, (c) represents Sigmoid response to clean signal, (d) is noisy input signal, (e) is ARSLIF response to noisy signal, (f) is Sigmoid response to noisy signal.}
	\label{fig-3}
\end{figure*}

The integration of Bi-LSTM and ARSLIF can be  described by the following mathematical equations: 
\begin{equation}
	\begin{aligned}
		f_t &= \text{ARSLIF}(W_f \cdot [h_{t-1}, x_t] + b_f) \\
		i_t &= \text{ARSLIF}(W_i \cdot [h_{t-1}, x_t] + b_i) \\
		\tilde{C}_t &= \tanh(W_C \cdot [h_{t-1}, x_t] + b_C) \\
		C_t &= f_t \times C_{t-1} + i_t \times \tilde{C}_t \\
		h_t &= \tanh(C_t) \times \text{ARSLIF}(W_o \cdot [h_{t-1}, x_t] + b_o)
	\end{aligned}
\end{equation}
where \(\text{ARSLIF}(\cdot)\) is used as the activation function.

Fig. \ref{fig-3} shows the heatmap visualization of noise robustness with the ARSLIF model. Figs. \ref{fig_3-1} and \ref{fig_3-4} depict the signals received by the model. The noisy input figure illustrates that the signal has been perturbed by Gaussian noise. Figs. \ref{fig_3-2} and \ref{fig_3-5} demonstrate that the ARSLIF model generates sparse pulse sequences, wherein the majority of time points yield zero output, with only a few time points emitting pulses. This characteristic is discernible from the color distribution in the heatmaps: most regions are colored in yellow (indicating absence of pulses), while only a few regions are in purple (indicating presence of pulses). Importantly, even in the presence of noise, the distribution pattern of pulses remains nearly unchanged, demonstrating robustness to noise interference. Figs. \ref{fig_3-3} and \ref{fig_3-6} implies that the Sigmoid function generates continuous values rather than sparse pulses. In the absence of noise, its output values span a wide range. However, in the presence of noise, there is a change in the distribution of output values, evident from the color variations in heatmaps. This implies that the output of the Sigmoid function is more susceptible to the influence of input noise.

The comparison between sigmoid and ARSLIF unveils a notable inspiration: even features with smaller values enable to initiate spikes, while a considerable proportion of features undergo deactivation. This phenomenon underlines the ARSLIF's robust regularization and effective noise filtering prowess. This also demonstrates that the enhanced model is particularly well-suited for addressing complex noise issues in audio-based depression diagnosis. By leveraging membrane potential leak mechanisms for effective noise filtration, accumulating membrane potential over time to capture temporal features, and enabling adaptive threshold adjustments, the model retains crucial information while eliminating redundancies and noise. The ARSLIF model introduces a flexible and powerful new approach to neural information processing. It emulates key characteristics of biological neurons and enhances adaptability to complex and heterogeneous signals. Specifically, in challenging tasks such as audio-based depression diagnosis, the model showcases significant potential and serves as one complement to the existing deep learning methods. 
\medskip

\section{Experiments and Results}\label{sec4}
To evaluate the model performance, experiments are conducted on the AVEC2014 dataset \cite{36} and on the MODMA dataset \cite{37}. The following subsections describe datasets, experimental setups and results, respectively.

\subsection{Datasets}
\textbf{The AVEC 2014 depression database} includes two tasks, namely "Northwind" and "FreeForm." In the "Northwind" task, participants were instructed to read a fable entitled "Die Sonne und der Wind" (English: "Northwind and Sun"). In the "FreeForm" task, participants were asked to answer several questions such as, "What is your favorite dish?" or to recount and describe a sad childhood memory in German. Across both tasks ("Northwind" and "FreeForm"), a total of 300 audio files were recorded, varying in duration from 6 seconds to 4 minutes. In other words, each task consists of 150 samples, which are then divided into training, development, and test sets, including 41, 43, and 42 speakers respectively. We merged the training, development, and test sets for both tasks, resulting in 100 audio samples in each set. In our experiments, we refer to this merged dataset as AVEC 2014, including training, validation and testing sets.

Considering potential biases of the original dataset, we resort to the mean Beck Depression Inventory (BDI) scores. In our examination of the timing for video recordings, there comes a notable observation: BDI scores of the same individual could undergo substantial fluctuations over brief durations. These fluctuations could even escalate from ``mild" to ``severe" within a matter of days. To mitigate this, we compute the mean BDI score across multiple videos for each participant within both the training and test sets.

We utilize the root mean square error (RMSE) and the mean square error (MAE) as model performance metrics. That is, \(\text{RMSE} = \sqrt{\frac{1}{K} \sum_{k=1}^{K} (y_k - \hat{y}_k)^2}\), where \( K \) represents the number of samples, \( y_k \) is the actual BDI-II score for the \( k^{th} \) sample, and \( \hat{y}_k \) is the predicted BDI-II score for the \( k^{th} \) sample. Define \(\text{MAE} = \frac{1}{K} \sum_{k=1}^{K} |y_k - \hat{y}_k|\), where \( K \) represents the number of samples, \( y_k \) is the actual BDI-II score for the \( k^{th} \) sample, and \( \hat{y}_k \) is the predicted BDI-II score for the \( k^{th} \) sample. The metrics \( y_k \) and \( \hat{y}_k \) denote the true and predicted BDI-II scores for the \( k^{th} \) sample, respectively. Smaller values for these metrics indicate a closer approximation to the actual scores, making them suitable for evaluating the performance of various algorithms.

\begin{table}[htbp]
	\renewcommand\arraystretch{1.3}
	\centering
	\caption{Optimal Training Parameters}
	\begin{tabular}{l||c}
		\hline
		\textbf{Training parameters} & \textbf{Setting} \\
		\hline
		ARSLIF Target Firing Rate ($f$) & 0.6 \\
		ARSLIF Initial Activation Threshold ($V_{\text{th}}$) & 0.8 \\
		ARSLIF active target rate ($f_{\text {active }}$) & 0.99 \\
		ARSLIF rest target rate ($f_{\text {rest }}$) & 0.01 \\
		ARSLIF threhold adaptation time constant ($\tau_{\text{adapt}}$) & 1000 \\
		Learning Rate & 0.0005 \\
		Batch Size & 96 \\
		\hline
	\end{tabular}
	\label{tab:op}
\end{table}

\subsection{Basic Parameters Setups}
The experiment starts with extracting audio from raw videos at a 22,050 Hz sample rate, employing the VideoFileClip library. To remove non-informative sections, a clipping threshold of -30dB is applied. The audio is then split into 32-frame windows and processed accordingly. Feature extraction, facilitated by the librosa library, yields MFCCs, first-order and second-order MFCC deltas\cite{25},\cite{39}, CQT\cite{42}, pitches\cite{24}, and jitter\cite{26} from each audio window. These features are normalized and combined to create multi-channel audio spectrograms.

To mitigate potential fluctuations in BDI scores, we employ an averaging approach. This reduces label inaccuracies by computing averages for each subject in both training and test sets. We organize data based on subject IDs and apply random cropping to ensure consistent audio lengths. Table \ref{tab:op} gives the training parameters, which are manually tuned by data characteristics and are determined through repeated tests. We use a V100 32GB platform and Keras version 2.9.0. Initial parameters in ARSLIF include a target firing rate of 0.6, an initial activation threshold of 0.8, with an adaptive target firing rate of 0.99 or 0.01. All the parameters are fixed during initialization and remain the same for all neurons at the single ARSLIF layer. The Adam optimizer with a learning rate of 0.0005 is used, while other parameters follow Keras defaults. We employ RMSE as the loss function and accelerate training with 16-bit mixed-precision. A batch size of 96 is chosen for efficiency. During testing, predictions are averaged per ID to generate the final BDI prediction. Model performance is evaluated through RMSE and MAE metrics, comparing predicted and actual scores. 

\subsection{Choosing Adaptive Threshold in Training}

\begin{figure}%[htbp]
  \centering
  \includegraphics[width=0.5\textwidth]{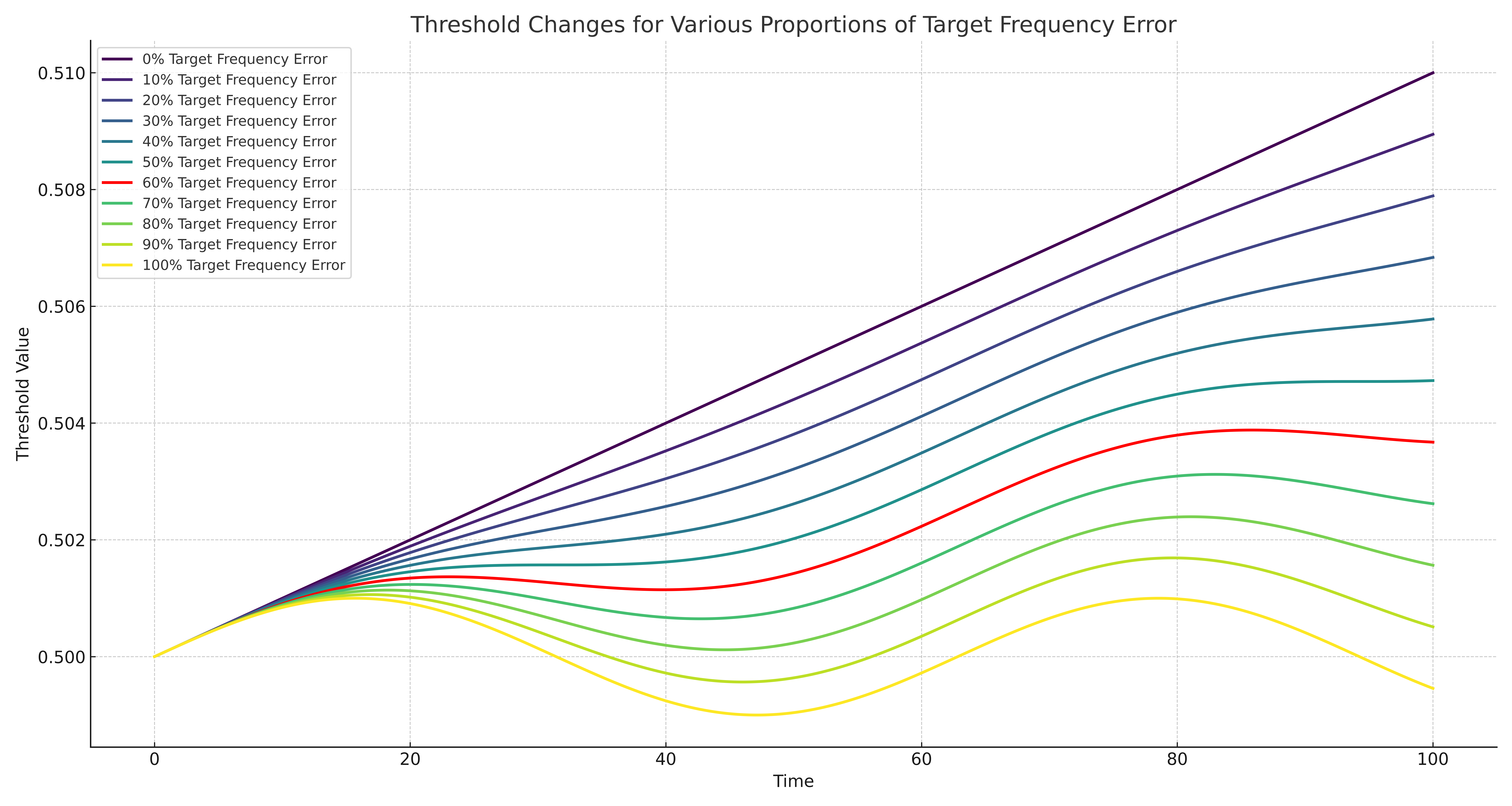}
  \caption{Changes of threshold under different target frequency error ratios $\frac{\alpha}{\beta+\alpha}$ as tested on the AVEC2014 dataset. 0\% target rate error ratio: The threshold continues to rise; 100\% target rate error ratio: The threshold oscillates; target rate error ratio ranging from 10\% to 90\%: The threshold oscillates slowly as it rises, with oscillation increasing as the error ratio increases.}
  \label{fig-4}
\end{figure}

Recalling the equation (\ref{eq-13}), at steady state, $V_{t h}(t)$ converges to a point where the rate of change is zero, the adjustment of the threshold \( V_{\text{th}} \) tends to make the firing rate \( r(t) \) approach \(\frac{\alpha f + \beta f_{\text{adapt}}(t)}{\alpha + \beta}\). Fig. \ref{fig-4} visualizes the threshold $V_{\rm th}$ trend under different target rate error ratios ($\frac{\alpha}{\alpha+\beta}$) as tested on the AVEC2014 dataset. Through experiments, we also proved that the larger the proportion of $\alpha$, the more inclined the threshold stability, and the larger the $\beta$, the more inclined the threshold rise, which is the same as the discussion conclusions in Subsection III.C.1). In the training process, as relevant features are filtered by the threshold, more significant parameters (weights) are retained. The gradual threshold increase reflects the process of feature selection and noise reduction. Here, we chose a 60\% target rate error ratio: $\alpha=0.6$ and $\beta=0.4$, which experimentally proves to effectively balance volatility and stability for AVEC2014.

\subsection{Comparison to The State-of-the-art Methods}
We conduct a comprehensive comparison of our model with other five methods on AVEC2014 dataset. Table \ref{tab:comparison} provides the detailed comparative results. The culmination of these results unequivocally illustrates performance achieved by RBE-FA in AVEC2014. 

\begin{table}[htbp]
	\renewcommand\arraystretch{1.3}
	\centering
	\caption{Comparison of RMSE and MAE across different methods.}
	\begin{tabular}{c||c||c}
		\hline
		Methods & RMSE & MAE \\
		\hline
		Valstar et al.\cite{37}  & 12.56 & 10.03 \\
		\
		Jain et al.\cite{40} & 11.51 & 9.74 \\
		
		Jan et al.\cite{41} & 10.28 & 8.07 \\
		
		He et al.\cite{28} & 9.99 & 8.19 \\
		
		Niu et al.\cite{43}& 9.66 & 8.02 \\
		
		RBA-FE (our) & 8.8310 & 8.8310 \\
		\hline
	\end{tabular}
	\label{tab:comparison}
\end{table}

RBE-FA has slightly weaker performance comparing with \cite{41}\cite{28}\cite{43} in MAE. Studies \cite{41} and \cite{28} showed better accuracy, especially in predicting lower scores on the Beck Depression Inventory (BDI). This improved accuracy might be attributed to the regression algorithms employed in these studies, which are seemingly more effective at estimating lower BDI scores. However, these studies also have higher Root Mean Square Error (RMSE) values, meaning they make bigger mistakes when predicting higher BDI scores. Conversely, RBE-FA has a lower RMSE, suggesting it's more accurate for higher BDI scores but might not be as precise for lower scores. In diagnosing depression, accurately identifying individuals with high BDI scores is particularly important to ensure that high-risk patients are not incorrectly diagnosed as low-risk.

Additionally, the model in \cite{43} uses significantly more features than RBE-FA, which could explain its higher accuracy but is less practical for home diagnosis due to its complexity. It is notable that the RBA-FE model employs a global max pooling step after a Temporal Convolutional Neural Network (TCNN), converting time-series data into a one-dimensional sequence. This process simplifies data but may lose critical intra-frame information that is vital for accurately predicting lower BDI scores, thus possibly contributing to a reduced MAE. One of our future work will focus on enhancing the model's ability to capture and utilize vital intra-frame information that is crucial for accurate prediction tasks.

\begin{figure*} [htbp]
\renewcommand\arraystretch{1.3}
\centering
\subfloat[]{\includegraphics[width=3in]{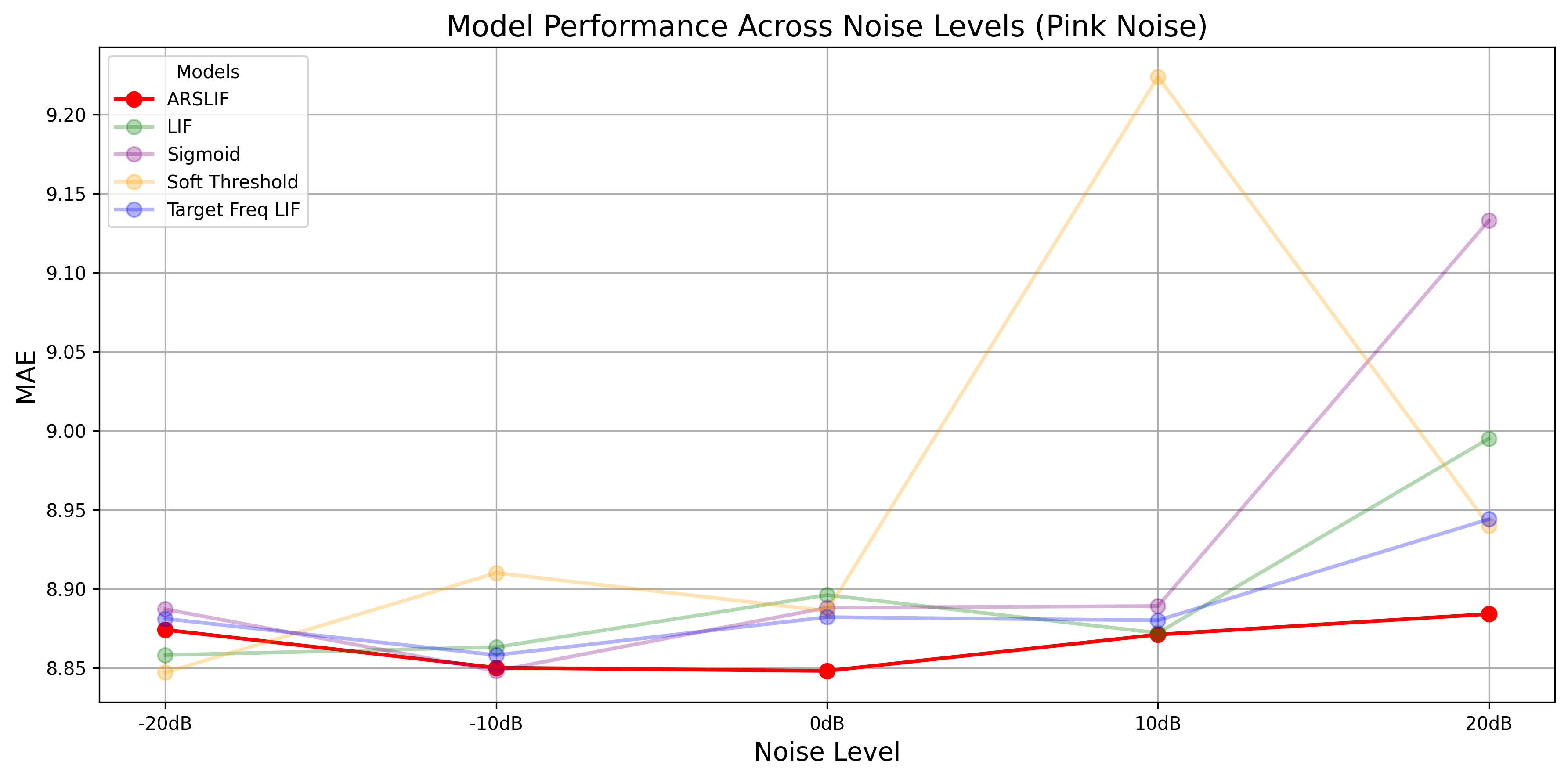}%
\label{fig_first_case}}
\hfil
\subfloat[]{\includegraphics[width=3in]{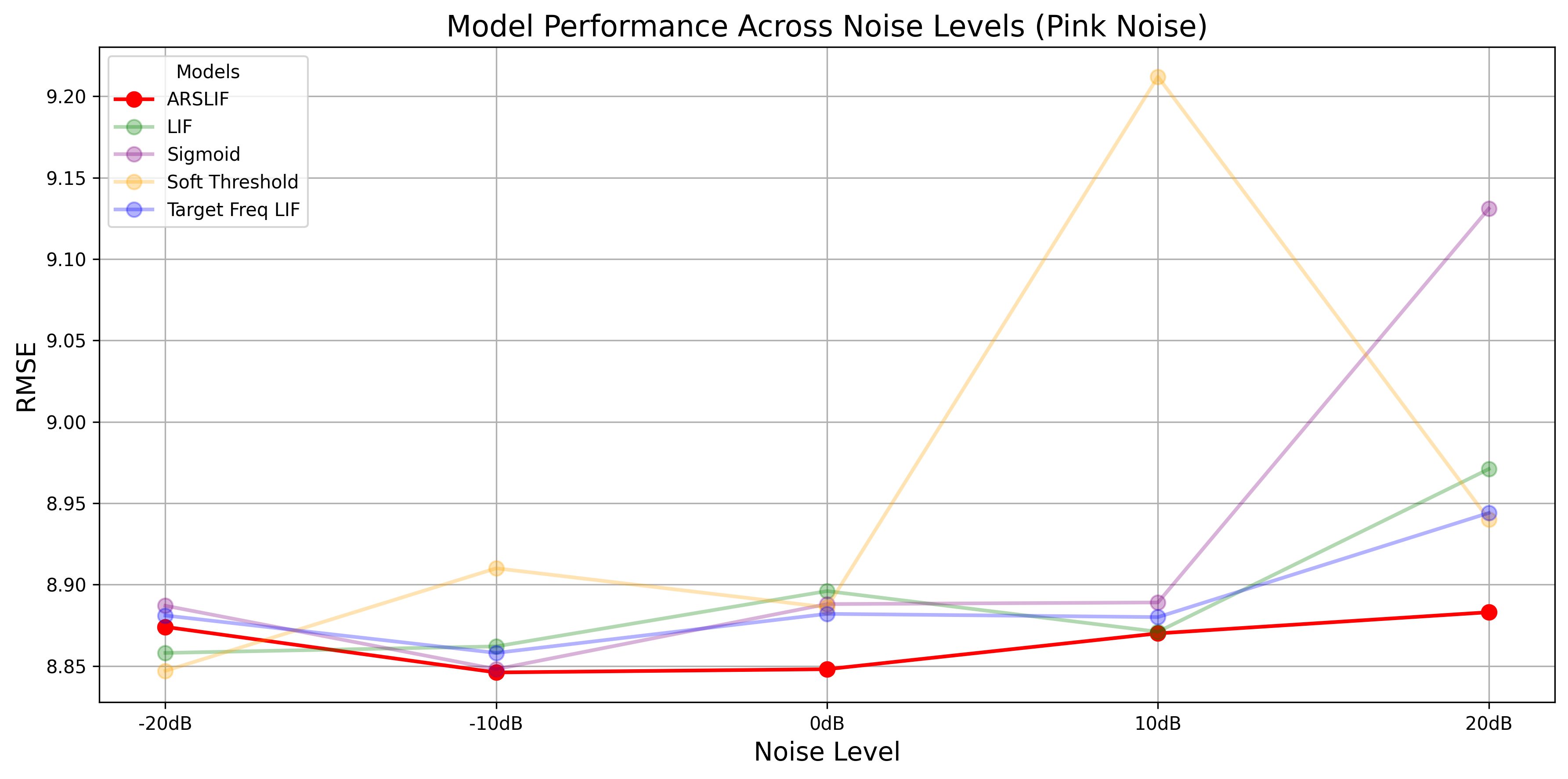}%
\label{fig_second_case}}
\hfil
\subfloat[]{\includegraphics[width=3in]{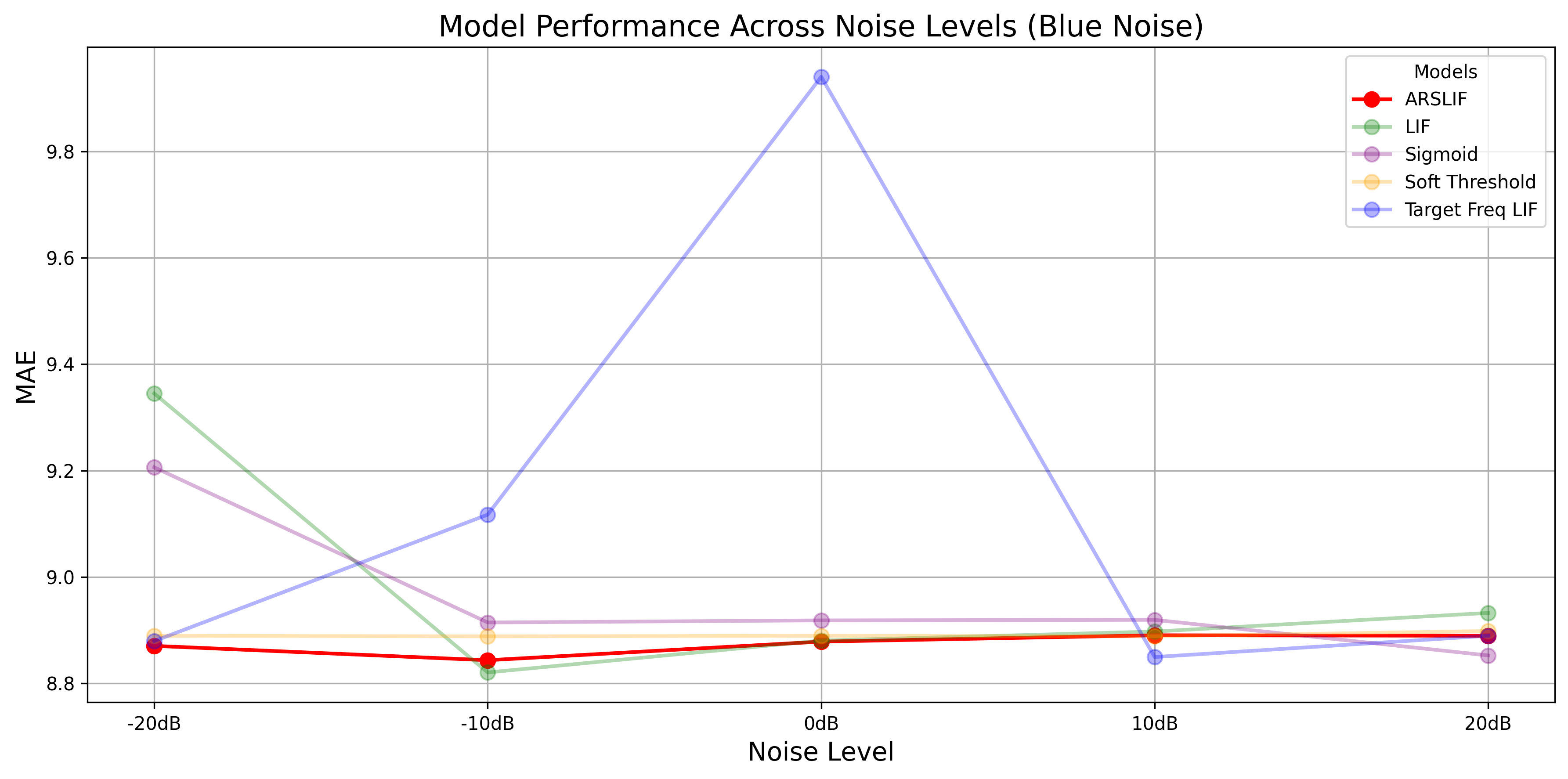}%
\label{fig_third_case}}
\hfil
\subfloat[]{\includegraphics[width=3in]{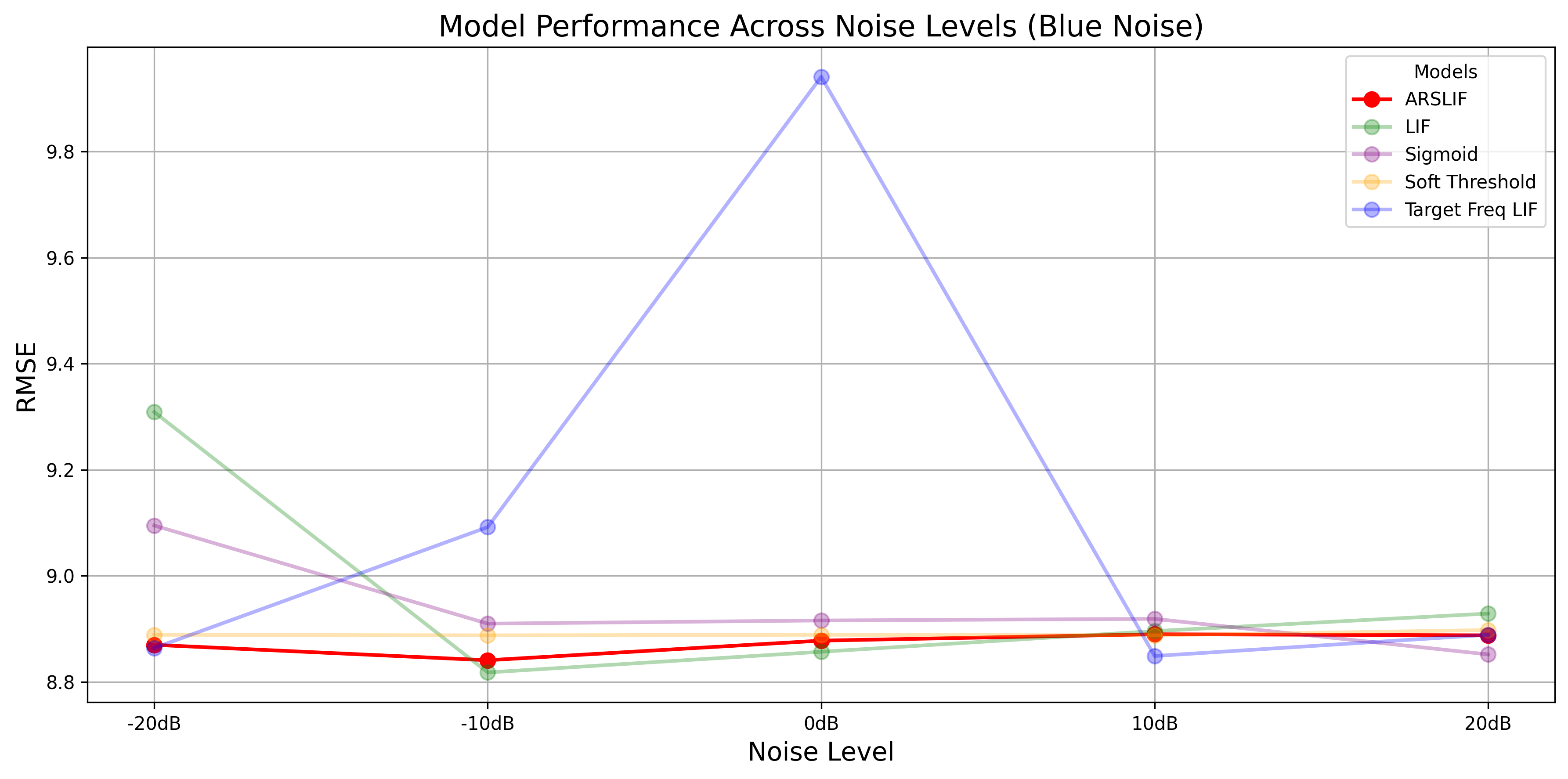}%
\label{fig_forth_case}}
\hfil
\subfloat[]{\includegraphics[width=3in]{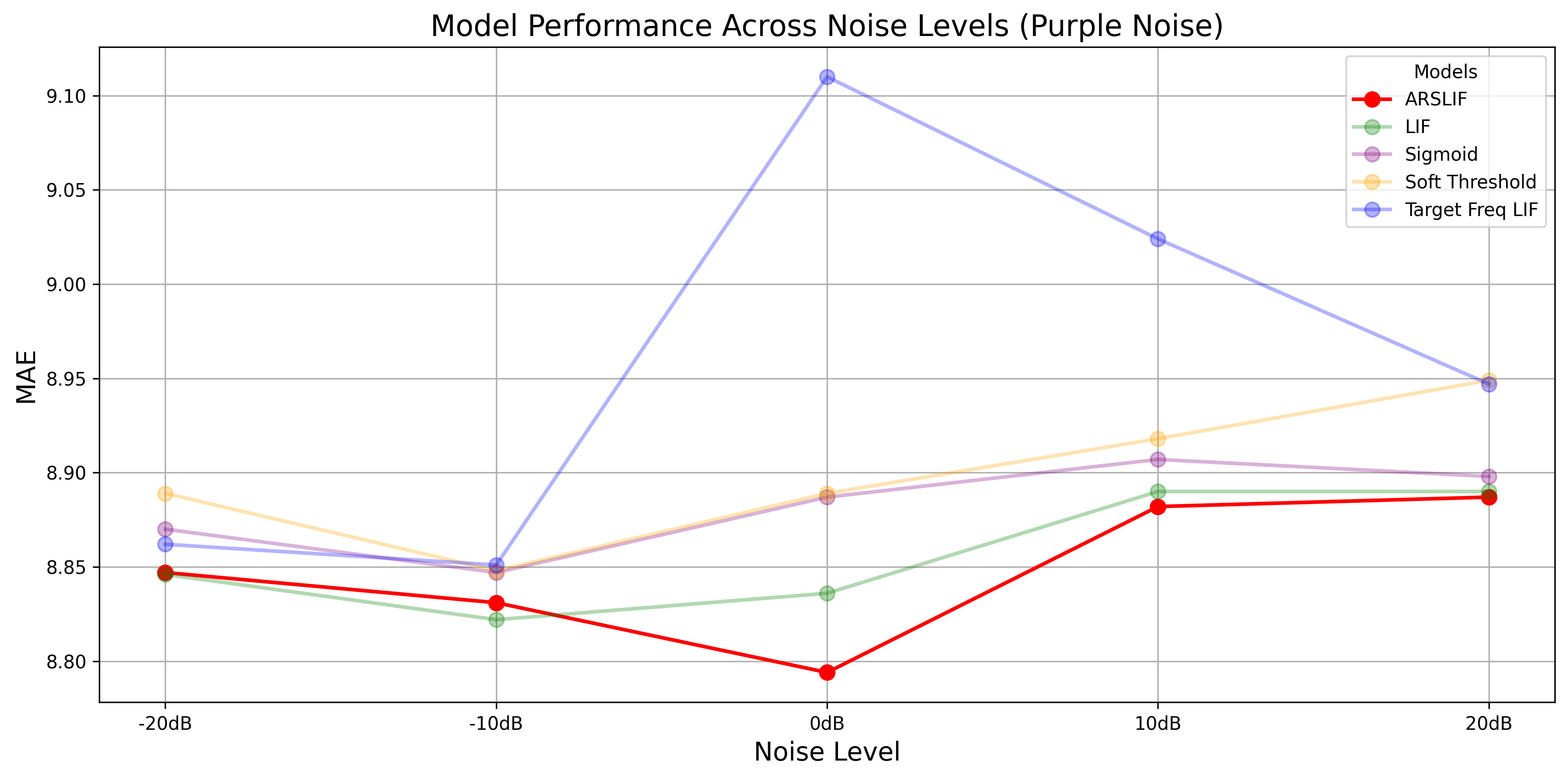}%
\label{fig_fifth_case}}
\hfil
\subfloat[]{\includegraphics[width=3in]{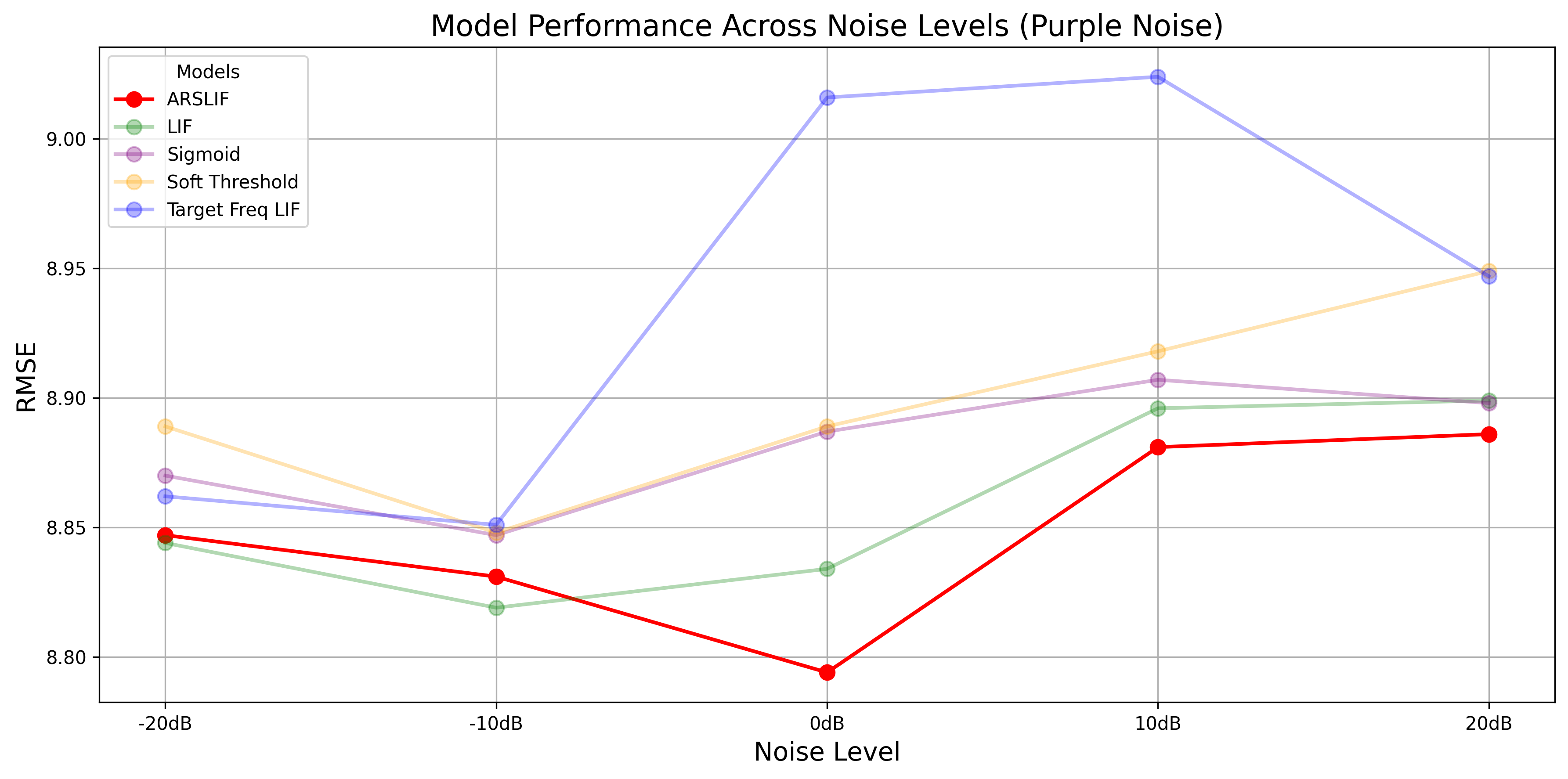}%
\label{fig_sixth_case}}
\caption{A comprehensive perspective for comparing the performance of four different activation functions against ARSLIF \textit{(in red)} via RMSE and MAE. In (a), (b), and (c), the model performance is assessed respectively across noise levels in terms of Root Mean Square Error (RMSE), the ARSLIF model consistently outperforms LIF, Sigmoid, Soft Threshold, and Target Freq LIF models, particularly at the extremes of the noise spectrum. Subfigures (d), (e), and (f) evaluate Mean Absolute Error (MAE), indicating the superior robustness of ARSLIF.}
\label{fig-5}
\end{figure*}

\subsection{Performance Evaluation of ARSLIF}
To validate the performance of ARSLIF, we incorporated four alternative gate activation functions—namely sigmoid, soft thresholding, standard LIF, and LIF with a desired firing rate $f$—into the RBA-FE model. Subsequently, we conducted a comparative analysis of their performance against ARSLIF. Besides, in order to assess the robustness of the model under real-world noise interference, this study incorporates three types of synthetic noise that are widely used : pink noise, blue noise, purple noise to emulate environmental noise into the dataset for performance testing. Specifically, pink noise is employed to simulate low-rate environmental noises generated by traffic and electrical appliances; blue and purple noises are utilized to mimic high-rate noises, such as those generated by human voices. The model's performance has been tested under varying levels of noise, ranging from -20dB to +20dB in relative volume. 

Fig. \ref{fig-5} presents a performance comparison of five gate activation functions in response to different noise levels, specifically blue, pink, and purple noise. Each graph provides insight into the model robustness and sensitivity by employing Root Mean Square Error (RMSE) and Mean Absolute Error (MAE) as the performance metrics.

\subsubsection{Compare ARSLIF with sigmoid}
we used the Sigmoid activation function as our baseline model and found that its effectiveness in detecting depression through speech decreased in noisy environments. It struggled with purple noise, showing increased error across all levels. For pink noise, its performance was stable until +10dB, where errors rose significantly. With blue noise, it matched other models until +10dB, then errors spiked, indicating a weakness to certain noise types at higher levels.

\subsubsection{Compare ARSLIF with soft thresholding}The soft thresholding is a commonly implemented noise attenuation technique within Deep Neural Networks (DNNs). In high noise levels, Soft threshold shows increased errors, especially under +20dB pink and 20dB purple noise. ARSLIF, despite some increased errors at +10dB in blue and pink noise, overall showed better stability and effectiveness in noisy environments.

\subsubsection{Compare ARSLIF with standard LIF}The LIF model reduced the impact of noise on diagnosing depression with a moderate performance with a gradual increase in RMSE as noise levels rise. But it struggled with adaptability due to its fixed threshold, leading to higher errors at increased noise levels. In contrast, ARSLIF shows a better robustness across different noise levels, maintaining lower error rates.

\subsubsection{Compare ARSLIF with `LIF with target firing rate'}
Here, we consider the target firing rate without considering the adaptive target firing rate, and equation (\ref{eq-7}) is changed into 
\begin{equation*}
	 r_e(t)= r(t)-f
\end{equation*}
where $r(t)$ represents the average rate at which all neurons fire at time $t$, $f$ is the target firing rate of neurons.

The LIF model with a target firing rate, while an improvement over the standard LIF, still did not match the adaptability of ARSLIF, especially under challenging purple noise conditions. ARSLIF demonstrated superior performance at +20dB noise level, maintaining lower error rates.
\smallskip

It is evident that in the presence of purple noise, the ARSLIF model exhibits superior and stable performance across noise levels, closely followed by soft thresholding. Although ARSLIF's performance at lower noise levels is not the best, with error rates peaking at the +10dB level for both blue and pink noise, considering the variability of noise levels in real-world environments, ARSLIF is deemed the most effective overall. In summary, ARSLIF could be considered the most reliable model overall among the five activation functions and have great robustness in both human and environment noise conditions.

\subsubsection{Compare ARSLIF with Other Adaptive Leaky Integrate-and-fire Models}
In order to further validate the advantage of ARSLIF, comparative work is done between ARSLIF and other adaptive leaky integrate-and-fire (LIF) models. Included are the Adaptive Exponential LIF (AdEx), Spike Response Model (SRM), and Multi-Timescale Adaptive Threshold Model. Table \ref{table:adaptiveLIF} demonstrates that ARSLIF achieves the highest F1 score and demonstrates the capacity to handling noisy audio data, among other adaptive LIF models.

\begin{table}[h]
	\renewcommand\arraystretch{1.3}
	\centering
	\caption{Comparison of F1 Scores between ARSLIF and other \newline adaptive LIF models} \label{table:adaptiveLIF}
\begin{tabular}{c||c}
	\hline
	Model & F1 Score \\
	\hline
	Adaptive Exponential (AdEx) LIF Model & 64.28\% \\
	Spike Response Model (SRM) & 62.04\% \\
	Multi-Timescale Adaptive Threshold Model & 62.04\% \\
	ARSLIF (Our) & 68.08\% \\
	\hline
\end{tabular} 
\end{table}

\smallskip

\subsection{Ablation Experiment}
An ablation experiment is performed on the AVEC2014 dataset, in order to validate the synergistic effect of multi-head attention and Bi-LSTM in the RBA-FE model. First, we remove the multi-head attention mechanism. It is shown that MAE has increased to 9.1419, and RMSE reaches 9.1998. Then, we eliminate the Bi-LSTM component, contributing to an MAE of 11.2401 and an RMSE of 11.2431. As illustrated in Table \ref{table:ablation_study}, the full RBA-FE model has an MAE of 8.8310 and an RMSE of 8.831, outperforming the two reduced counterparts. The focus of such an ablation experiment is to assess the contribution of each component to the proposed RBA-FE model when used for extracting audio features with depression. Table \ref{table:ablation_study} suggests that the removal of either component adversely impacted the feature extraction capabilities of the hierarchical feature extractor, underscoring that each network structure within the hierarchical classifier plays a crucial role in the effective extraction of speech features for depression diagnosis. Through these ablation experiments, it has been substantiated that both the attention and Bi-LSTM mechanisms contribute to the overall robustness and performance of the RBA-FE model, and Bi-LSTM can be seen as the core component of RBA-FE model.

\begin{table}[htbp]
	\renewcommand\arraystretch{1.3}
	\centering
	\caption{Ablation Experiment on AVEC2014 Dataset}
	\label{table:ablation_study}
	\begin{tabular}{c||c||c}
		\hline
		Model Variant & MAE & RMSE \\
		\hline 
		Full Model (RBA-FE) & 8.8310 & 8.8310 \\
		Without Multi-Head Attention & 9.1419 & 9.1998 \\
		Without Bi-LSTM & 11.2401 & 11.2431 \\
		\hline
	\end{tabular}
\end{table}

\subsection{Generalization Experiment}
To demonstrate the generalization of the RBA-FE model, a similar experiment is conducted on another depression audio dataset: MODMA \cite{38}. Additionally, we compare the RBA-FE model with other state-of-the-art models on the DAIC-WOZ dataset \cite{53} to justify the improvements.

\subsubsection{Experiment on MODMA dataset}
The MODMA audio dataset originates from the Second Hospital of Lanzhou University in China and is a carefully curated experimental dataset that encompasses both depression diagnosis and healthy controls. The dataset consists of 52 participants, with 23 diagnosed with depression and 29 as healthy controls. Each participant was meticulously selected based on explicit inclusion and exclusion criteria and was recruited under rigorous ethical guidelines. The depression group included 23 outpatients (16 males and 7 females, ages 16-56), diagnosed by clinical psychiatrists and evaluated using diagnostic tools such as DSM-IV and PHQ-9. The healthy control group consisted of 29 healthy participants (20 males and 9 females, ages 18-55). Exclusion criteria encompassed psychiatric disorders, brain organ injuries, severe physical illnesses, and severe suicidal tendencies.

The experiment has three core components, lasting approximately 25 minutes. It included 18 interview questions targeting positive, neutral, and negative emotions; reading two sets of common Chinese words and a short story; and describing three facial expression images and one emotional theme image. The entire experiment was conducted in Mandarin Chinese, adopting a randomized sequence to offset any sequence effects. The recording equipment comprises a Neumann TLM102 microphone and an RME FIREFACE UCX audio interface, with a sampling rate of 44.1 kHz and a bit depth of 24 bits. The experimental setup is a soundproof room free of electromagnetic interference. On the data recording and technical validation front, all records were saved in uncompressed WAV format, and the recording quality was quantitatively assessed via the signal-to-noise ratio (SNR), ranging from 20 to 30.

\smallskip

\begin{figure}%[ht]
	\centering
	\includegraphics[width=0.49\textwidth]{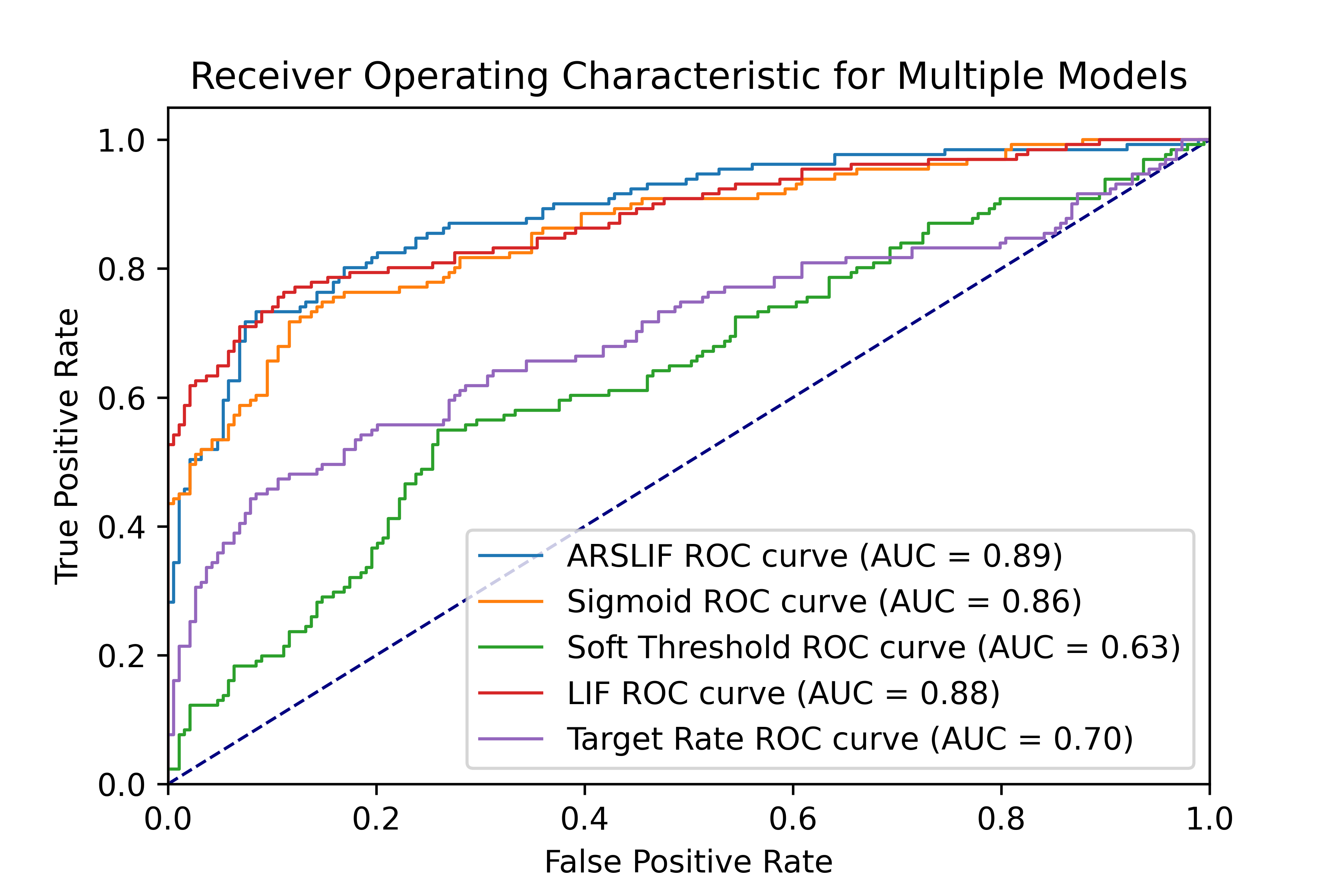}
	\caption{ROC curves of comparative models on the testing MODMA dataset}
	\label{fig-6}
\end{figure}

Since MODMA is a binary-labeled dataset, no additional label manipulations have to be considered. We allocate 80\% of all audio IDs in this dataset for training, while the remaining 20\% are for testing. Stratified sampling was employed for data division to ensure equal label distribution between the training and test sets. Given the binary nature of the labels, data segments were divided into lengths of 128. For the model training, the following parameter settings are employed: the target firing rate for ARSLIF was set to 0.6, the initial activation threshold was set at 0.6. The optimizer of choice was Adam with a learning rate of 0.001, while all other parameters were kept at their default settings. The loss function utilized was categorical cross-entropy. During the training phase, we configured the batch size to be 96 and leveraged 16-bit floating-point mixed precision to accelerate the training process. In the testing phase, inferences were made on all audio segments, and the final prediction for each audio clip was determined using a hard voting scheme.

\begin{table}[htbp]
\renewcommand\arraystretch{1.3}
\centering
\caption{Performance comparison of different methods on MODMA dataset}
\label{tab:performance_comparison}
\begin{tabular}{c||c||c||c||c}
\hline
Methods & Precision & Accuracy & Recall & F1 Score \\
\hline
DenseNet \cite{44} & 0.8318 & 0.8446 & 0.8652 & 0.8308 \\

EFNerV1 \cite{45} & 0.8168 & 0.7944 & 0.8177 & 0.7924 \\

Effnetv2m \cite{46} & 0.8268 & 0.6999 & 0.8014 & 0.7912 \\

Efnetv2s \cite{46} & 0.8192 & 0.6974 & 0.7930 & 0.7222 \\

SENet \cite{27} & 0.8110 & 0.7698 & 0.7129 & 0.7846 \\

RBA-FE (Our) & 0.8750 & 0.8974 & 0.8750 & 0.8750 \\
\hline
\end{tabular}
\end{table}

Table \ref{tab:performance_comparison} shows the classification performance of the proposed RBA-FE model, with comparison to related methods. It is demonstrates that our model achieves an accuracy of 89.74\% on the testing set, outperforming other models specialized in depression audio recognition on MODMA. This validates the efficacy of our model and parameter configurations. 
\smallskip

Fig. \ref{fig-6} further depicts the corresponding ROC (Receiver Operating Characteristic) curves. The blue line represents the performance of a random classifier. For a binary classification problem, it has a 50\% chance of predicting a sample as the positive class and a 50\% chance of predicting it as the negative class. Consequently, this line forms a 45-degree diagonal from the origin (0,0) to the point (1,1). The orange curve represents the ROC curve of our model. This curve illustrates the relationship between the True Positive Rate (TPR) and the False Positive Rate (FPR) at various threshold levels. The closer the curve is to the upper left corner, the better the performance of the model. The area under curve (AUC) is a quantifiable metric used to gauge the overall performance of the model. AUC values range between 0 and 1; a perfect classifier would have an AUC of 1, whereas a random classifier would have an AUC of 0.5. Our model has an AUC of 0.89, indicating a high level of classification performance.
\smallskip

\subsubsection{Experiment on the DAIC-WOZ dataset}
The DAIC-WOZ audio dataset originates from the Distress Analysis Interview Corpus, which is widely used for depression detection and other mental health-related studies. This dataset includes audio and text from clinical interviews conducted with participants, aiming to distinguish between individuals diagnosed with depression and healthy controls. The DAIC-WOZ dataset consists of interviews that were part of a Wizard-of-Oz experiment, where a human interviewer (wizard) controlled the virtual agent that interacted with the participant. The dataset contains both audio and transcript data, with participants’ depression severity being assessed using diagnostic tools such as PHQ-9.

In this study, no additional label manipulations were applied to the dataset. We use the official split of training and test sets based on the provided IDs. The training set includes participants with both depression and non-depression labels, while the test set is used to evaluate the generalization of the model.

For model training, the target firing rate for ARSLIF was set to 0.6, and the initial activation threshold was also set at 0.6. The Adam optimizer with a learning rate of 0.001 was used, and categorical cross-entropy served as the loss function. The batch size was set to 96 to enable efficient processing during training, which was further accelerated by leveraging 16-bit floating-point mixed precision without compromising accuracy. During the testing phase, inferences were made on all audio segments in the test set, and a hard voting scheme was employed to determine the final prediction for each audio clip. These predictions were then compared with the ground truth to calculate the model performance on the DAIC-WOZ dataset.

\begin{table}[htbp]
	\renewcommand\arraystretch{1.3}
	\centering
	\caption{Performance comparison of different methods on DAIC-WOZ dataset}
	\label{tab:performance_comparison2}
	\begin{tabular}{c||c}
		\hline
		Methods & F1 Score \\
		\hline
		DEPA \cite{54} & 0.6400 \\
		
		SIDD \cite{55} & 0.6010 \\
		
		DepAudioNet \cite{56} & 0.3800 \\
		
		FRAUG \cite{57} & 0.4790 \\
		
		RBA-FE (Our) & 0.6808 \\
		\hline
	\end{tabular}
\end{table}

Table \ref{tab:performance_comparison2} presents the classification performance of the proposed RBA-FE model compared to other related methods. To ensure a fair and direct comparison, we focused solely on the audio-only versions of the models, where feature extraction is done at the frame level, and no pretrained models were used. Our RBA-FE model achieved an F1 score of 68.08\% on the DAIC-WOZ test set, outperforming other baseline models specialized in depression audio recognition. This highlights the effectiveness of our model and the selected parameter configurations in managing depression audio datasets.

\section{Conclusion}\label{sec6}
In this article, a brain-inspired robust model has been developed for depression audio diagnosis. This RBA-FE model intergrades a temporal convolutional neural network, multi-head attention, and bidirectional long short-term memory networks, and leverages the unique strengths of each element to tackle complex diagnostic challenges. This article further presents an ARSLIF neuron model, which mimics the ``cell selectivity" in the brain attention systems, which can deal with the gate overfitting and noise sensitivity in LSTM by using an adaptive threshold. Experiments on three public datasets: AVEC2014, MODMA, DAIC-WOZ, have substantiated the effectiveness and generalization of the RBA-FE model. 

The comparative analyses further support that the ARSLIF neurons excel in capturing subtle features in noisy environments. It is demonstrated that stricter thresholds and target frequencies, as well as higher activation frequencies, yield better robustness. This observations fills a research gap in the interdisciplinary domain intersecting machine learning-based depression audio diagnosis and neuroscience, and for the first time we introduce the``cell selectivity" into LIF neurons. This provides an efficacious solution for audio diagnosis of depression and opens new avenues for future research in this field. Future work will focus on investigating the operational mechanisms of the ARSLIF neuron, including its specific filtering roles and adaptive processes within the network. This may involve visualizing and analyzing the internal neuronal activity within the network to better understand the model outputs, e.g., in abnormality detection or  landmark localization tasks \cite{11,47}. Overall, this study not only advances the technology for audio diagnosis of depression but also provides robust support for interdisciplinary research in this area.

\end{document}